\documentclass[superscriptaddress, twocolumn,showkeys,
amssymb,amsmath,nobibnotes,aps,prd,
nofootinbib]{revtex4-2}
\pdfoutput=1
\usepackage{graphicx,subfigure,bm,color,psfrag,hyperref}
\usepackage[export]{adjustbox}
\usepackage{amsfonts}
\usepackage{lipsum}
\usepackage{mathtools}
\usepackage{verbatim}
\usepackage[normalem]{ulem}
\usepackage[dvipsnames]{xcolor}
\hypersetup{colorlinks,linkcolor={blue},citecolor={red},urlcolor={magenta}}

\begin{document}

\title{Interacting phantom dark energy: new accelerating scaling attractors}

\author{Sudip Halder}
\email{sudip.rs@presiuniv.ac.in}
\affiliation{Department of Mathematics, Presidency University,  86/1 College 
Street, Kolkata 700073, India}

\author{S.~D.~Odintsov}
\email{odintsov@ice.csic.es}
\affiliation{ICREA, Passeig Luis Companys, 23, 08010 Barcelona, Spain}
\affiliation{Institute of Space Sciences (ICE, CSIC), Carrer Can Magrans, s/n,  08193 Barcelona, Spain}

\author{Supriya Pan}
\email{supriya.maths@presiuniv.ac.in}
\affiliation{Department of Mathematics, Presidency University, 86/1 College  
Street, Kolkata 700073, India}
\affiliation{Institute of Systems Science, Durban University of Technology,  PO 
Box 1334, Durban 4000, Republic of South Africa}

\author{Tapan Saha}
\email{tapan.maths@presiuniv.ac.in}
\affiliation{Department of Mathematics, Presidency University, 86/1 College 
Street,  Kolkata 700073, India}

\author{Emmanuel N. Saridakis}
\email{msaridak@noa.gr}
\affiliation{National Observatory of Athens, Lofos Nymfon,  11852 Athens, 
Greece}
\affiliation{CAS Key Laboratory for Research in Galaxies and  Cosmology, 
Department of Astronomy, School of Physical Sciences, University of Science and 
Technology of China, Hefei, Anhui 230026, China}
\affiliation{Departamento de Matem\'{a}ticas, Universidad  Cat\'{o}lica del 
Norte, Avenida Angamos 0610, Casilla 1280 Antofagasta, Chile}


\begin{abstract}

We perform a detailed investigation of interacting phantom cosmology, by 
applying the powerful method of dynamical system analysis. We consider two 
well-studied interaction forms, namely one global and one local one, while the 
novel ingredient of our work is the examination of new potentials for the 
phantom field. Our analysis shows the existence of saddle 
matter-dominated points, stable  dark-energy dominated points,  and scaling 
accelerating solutions,  that can attract the Universe at late times. 
As we show, some of the stable accelerating scaling attractors, in 
which dark matter and dark energy can co-exist, alleviating the cosmic 
coincidence problem, are totally new, even for the previously studied 
interaction rates,  and arise purely from the novel potential forms.   
\end{abstract}


\maketitle

\section{Introduction}
\label{sec-introduction}

Dark matter (DM) and dark energy  (DE) are two main ingredients of the 
universe, constituting of  95\% of its total energy budget. 
The DM sector is responsible for the observed structure formation, while  
the DE sector is in charge of the late time accelerated expansion  
\cite{Copeland:2006wr,Bamba:2012cp}. The nature, origin and dynamics of both DE 
and DM are not yet clearly understood, nevertheless    the $\Lambda$-Cold Dark 
Matter ($\Lambda$CDM) cosmological model, based on General 
Relativity, matches quite well with a large span of the astronomical datasets 
in 
which DE is a simple cosmological constant and DM is a cold pressureless 
sector. However, $\Lambda$CDM faces many challenges at both theoretical and 
observational levels 
\cite{DiValentino:2021izs,Perivolaropoulos:2021jda,Abdalla:2022yfr}, and among 
possible improvements  is the consideration of a time varying nature in the DE 
sector. 

In general, DE is assumed to evolve independently and  interact only with  
gravity. However, there is no fundamental reason to exclude   possible 
interactions between  DE and other fields, especially with the DM sector whose 
microscopic nature is also unknown. The interaction between DE and DM can 
alleviate the cosmic coincidence problem 
\cite{Amendola:1999er,Chimento:2003iea,Cai:2004dk,Pavon:2005yx,Huey:2004qv, 
Hu:2006ar,delCampo:2008sr,delCampo:2008jx}, while  it can potentially  
alleviate the current cosmological tensions too
\cite{DiValentino:2021izs,Perivolaropoulos:2021jda,Abdalla:2022yfr,
Kumar:2017dnp,DiValentino:2017iww,Yang:2018euj,Kumar:2019wfs,Pan:2019jqh, 
Pan:2019gop,DiValentino:2019ffd,Pan:2020bur, Pan:2023mie}. 
Hence, over the last years interacting DM-DE scenarios have gained 
significant attention in the literature for their attractive features  
\cite{Guo:2004xx, Sussman:2005mf,Zimdahl:2005bk,Wang:2005jx,
Wang:2005ph, Barrow:2006hia,Setare:2006wh, Poplawski:2006kv,
Sadjadi:2006qb,Chang:2006dj,Kim:2006kk,Rosenfeld:2007ri,Zimdahl:2007zz,
Feng:2007wn, Cruz:2008qp, 
 Chen:2008ca,
Jamil:2009eb,   Jackson:2009mz,Valiviita:2009nu, 
Suwa:2009gm,Karami:2009yd,Wei:2010uh,
 Martinelli:2010rt,Gavela:2010tm, 
LopezHonorez:2010esq,Baldi:2010pq, Chen:2011cy,Baldi:2011wy, 
Chimento:2013rya,Harko:2012za,Sun:2013pda,Li:2013bya,Tamanini:2013xia,
Yang:2014gza,   
Li:2014cee,Pan:2012ki,Duniya:2015nva,
Li:2015vla,Odderskov:2015fba,Nunes:2016dlj, Ferreira:2014jhn,
Biswas:2016idx,
Costa:2016tpb, Linton:2017ged, Yang:2018ubt, 
 DiValentino:2019jae,
Li:2019loh, 
Yang:2020uga,Pan:2020zza,Pan:2020mst,
DiValentino:2020kpf,Yang:2021hxg,Gao:2021xnk,Chatzidakis:2022mpf,Hou:2022rvk,
Pan:2022qrr,Yang:2022csz, 
Zhai:2023yny,Bernui:2023byc,Escamilla:2023shf,Benisty:2024lmj,Halder:2024uao,
Giare:2024ytc,Giare:2024smz,Aboubrahim:2024spa,Sabogal:2024yha,Ghedini:2024mdu,
Paliathanasis:2024abl,Aboubrahim:2024cyk} (for reviews 
 see \cite{Bolotin:2013jpa,Wang:2016lxa}).

The interacting DM-DE models, widely known as Interacting DE (IDE) or Coupled 
DE, induce a richer dynamics in the universe evolution, recovering the 
non-interacting scenarios as special cases. The heart of the IDE models is the 
interaction function which characterizes the rate of energy and/or momentum 
transfer between the dark components, and consequently it affects the 
evolution of the universe.
In order  to investigate the dynamics of   interacting cosmologies, one can 
  confront them with   observational datasets from various astronomical 
surveys and extract constraints on the involved functions and parameters. 
However,    one can also apply the powerful approach of dynamical 
systems, which allows  to extract   analytical 
information independently of the initial conditions. 
The techniques of dynamical system analysis  have been widely applied in 
the context  of non-interacting and interacting dynamics 
\cite{Copeland:1997et,Billyard:2000bh,Gong:2006sp,Chen:2008pz,Chen:2008ft,
Boehmer:2008av,Karwan:2008ig,Caldera-Cabral:2008yyo,Leon:2009ce,Leon:2009dt,
Xu:2012jf,Leon:2012mt,Avelino:2013wea,Boehmer:2015kta,Boehmer:2015sha,
Biswas:2015zka,Fadragas:2013ina,Biswas:2015cva,Mahata:2015nga,Banerjee:2015kva,
Paliathanasis:2015gga,Paliathanasis:2015arj,Singh:2015rqa,Dutta:2017kch,
Carneiro:2017evm,Odintsov:2017icc,Zonunmawia:2017ofc,Bahamonde:2017ize,
Odintsov:2018uaw,Paliathanasis:2019hbi,Hernandez-Almada:2020ulm,Sa:2020fvn,
Chakraborty:2020yfe,Paliathanasis:2021egx,Sa:2021eft,Samart:2021viu,
Arevalo:2022sne,Hussain:2023kwk,Sa:2023coi,Saha:2024xbg,Halder:2024gag}.

In the present work we are interested in performing a dynamical-system analysis 
of interacting cosmology  considering a phantom dark energy \cite{Caldwell:1999ew,Dabrowski:2003jm,Singh:2003vx,Nojiri:2005sx,Arefeva:2005ixb,
Capozziello:2005tf,Saridakis:2008fy,Cai:2009zp}. 
In particular, we consider various interacting phantom DE scenarios, 
characterized by four different potentials and  two 
specific interacting functions, one corresponding to the global interaction 
rate and the other corresponding to the local interaction rate.  
The examination of new potentials even for the known and studied cases of 
interaction form reveals new 
accelerating scaling attractors that  were never studied in the 
literature. This is an important result, since it was  believed 
that in general, due to 
 the phantom nature of the acceleration, accelerating scaling solutions in 
phantom cosmology were very rare.

The manuscript is structured as follows. 
In section \ref{sec-basic-eqns} we present the basic equations   of  general 
IDE 
scenarios and we construct the various specific models. In 
section~\ref{sec-dyn-systems} we perform the detailed phase space analysis 
extracting all possible solutions. 
  Finally,   section~\ref{sec-conclusion}
is devoted to the conclusions.

\section{Phantom Interacting Dark Energy }
\label{sec-basic-eqns}

In this section we briefly present the scenario of phantom interacting dark 
energy, providing the main equations. 
We consider a flat  Friedmann-Lema\^{i}tre-Robertson-Walker (FLRW) line element 
given by 
\begin{eqnarray}\label{flrw}
 ds^2 = -dt^2 + a^2(t) \left[  dr^2 + r^2 (d\theta^2 + \sin^2 \theta 
d\phi^2) \right],   
\end{eqnarray}
where $a(t)$ is the expansion scale factor of the universe.
Concerning the gravitational theory we consider   standard general 
relativity, and we additionally consider the DM
sector of the universe corresponding to a 
perfect fluid. Concerning the dark energy sector, we attribute it to a phantom 
scalar field  $\phi$ with action 
\begin{eqnarray}
    S_{\phi} = \int d^4 x \sqrt{-g} \left[ \frac{1}{2} (\nabla \phi)^2 -  V 
(\phi) \right],
\end{eqnarray}
where $V (\phi)$ is the potential. In the case of FLRW geometry, the energy 
density  $\rho_{\phi}$  and 
pressure $p_{\phi}$ of the phantom fluid are given by 
\begin{eqnarray}
	&& \rho_{\phi} = - 
\frac{\dot{\phi}^2}{2} + V (\phi),\label{rhoph1}\\
	&& p_{\phi} = -\frac{\dot{\phi}^2}{2} - V (\phi),\label{pphi2}
\end{eqnarray}
where a dot denotes derivative with respect to the cosmic time. 
Additionally, the  Friedmann equations   are written as
\begin{eqnarray}
	&&  3H^2 =\kappa^2(\rho_m+\rho_\phi  )  \label{friedmann-1}\\
	&& 2\dot{H}+3H^2= -\kappa^2(p_m+p_\phi)  ,\label{friedmann-2}
\end{eqnarray}
where  $H = \dot{a}/a$ is  the Hubble function (dots denote time-derivatives) 
  $\kappa^2= 8\pi G$ is the  gravitational constant,
and   $\rho_m$, $p_m$ are respectively the energy density and pressure of the DM 
fluid.

We mention here that 
  in phantom cosmology the kinetic term is negative,  however the sign of the 
total energy of the phantom field  depends on its
potential too. Hence, if one chooses potentials for which
  $V(\phi)> \dot{\phi}^2/2$ holds, then $\rho_{\phi}$ 
remains positive. 
However, even 
if the total phantom energy  becomes negative, it does not   lead to 
pathologies, since   the matter energy density in (\ref{friedmann-1}) 
compensates this situation and 
thus  $H^2$ remains always positive, as it should be. In other words, the fact 
that in phantom cosmology one can always impose a specific potential, and/or 
interaction form, that will make everything well-behaved and self-consistent 
(this is the case in many modified gravity theories, e.g. Horndeski 
gravity \cite{DeFelice:2011bh}, where in principle they can lead to negative 
$H^2$, however one 
chooses the involved functions in a suitable way in order to always have 
$H^2>0$).

 In summary,   at the classical level the 
negativity of $\rho_{\phi}$ is not a problem, and
all energy conditions are the same with the case of canonical fields.  
Nevertheless, issues might arise concerning the quantum field theory 
description of 
phantom fields, since   unbounded-from-below states may lead to problems. 
However, in the 
literature there have   been various attempts in  constructing a phantom 
theory consistent with the basic requirements of quantum field theory, for 
example considering that 
the  phantom fields arise as   effective descriptions of more fundamental 
canonical theories \cite{Nojiri:2003vn,
Nojiri:2003ag}.

The key idea of interacting cosmology is that although the total energy 
content of the universe is conserved (as a requirement of the diffeomorphism 
invariance) the separate DE and DM sectors are not conserved, but mutually 
interact. In particular, their conservation equations  take the general form
\begin{eqnarray}
&& \dot{\rho}_m + 3 H (1+w_m) \rho_m = -Q ,\label{cons-CDM}\\
&& \dot{\rho}_\phi + 3 H (1+w_{\phi}) \rho_\phi =  Q ,\label{cons-DE}
\end{eqnarray}
where $w_m = p_m/\rho_m$, $w_{\phi} = p_{\phi}/\rho_{\phi}$ are respectively the barotropic equation of state parameter of DM and DE, and  
$Q$ is  the interaction function which characterizes the transfer of 
energy between the dark sectors. For  $Q > 0$, the energy flow occurs from DM 
to DE, while for $Q <0$  energy flows in the reverse 
direction, namely from DE to DM. Notice that eqn. 
(\ref{cons-DE}) can alternatively be expressed in a Klein-Gordon-like form, 
namely 
\begin{eqnarray}
    \Ddot{\phi}+3H\dot{\phi}- \frac{\partial 
V(\phi)}{\partial\phi}=-\frac{Q}{\dot{\phi}}.
\end{eqnarray}
In this way, one can see that if the interaction function $Q (t)$ and the 
potential $V (\phi)$ of the scalar field are prescribed, then the evolution of 
the interacting fluids can be determined in principle. 

In summary,  the choice of both functions  $Q (t)$ and  $V (\phi)$ is the 
starting point for any analysis. In particular, using the 
conservation equations, one can find the evolution laws of the dark fluids as 
 \begin{eqnarray}
     \rho_{m} = \rho_{m,0} e^{\phi(1) - \phi(a)} - e^{-\phi(a)} \int_{1}^{a} \frac{Q e^{\phi(a)}}{aH} da,\\
     \rho_{\phi} = \rho_{\phi,0} e^{\psi(1) - \psi(a)} + e^{-\psi(a)} \int_{1}^{a} \frac{Q e^{\psi(a)}}{aH} da,
 \end{eqnarray}
where $\phi(a) = 3 \int a^{-1}(1+w_m)~ da$, $\psi (a) = 3 \int a^{-1} 
(1+w_{\phi})~da$, and where 
$\rho_{m,0}$, $\rho_{\phi, 0}$ are respectively the present-day  value of 
$\rho_{m}$, $\rho_{\phi}$. The above equations show that  when a mechanism of 
energy transfer is allowed between the dark fluids, the interaction rate 
affects their evolution.  Furthermore, in presence of such interaction,  the
effective behavior of the phantom DE could be quintessential.  This  can become 
transparent by rewriting the conservation equations 
(\ref{cons-CDM}), (\ref{cons-DE}) as  
$\dot{\rho}_m + 3 H (1+w_{m}^{\rm eff}) \rho_m =0$ and $\dot{\rho}_{\phi} + 3 H 
(1+w_{\phi}^{\rm eff}) \rho_{\phi} = 0$, where  $w_{m}^{\rm eff} = w_m + Q/(3H 
\rho_m)$ and $w_{\phi}^{\rm eff} = w_{\phi} - Q/(3H \rho_{\phi})$  are 
respectively the effective equation-of-state parameter for dark matter   and 
dark energy. Thus, one can see that even if $w_{\phi} < -1$, $w_{\phi}^{\rm 
eff}$ could be larger than $-1$.

In the following we focus on two interacting functions. The first interaction 
model is based on the most well-studied interaction form,  namely
\cite{Bolotin:2013jpa,Wang:2016lxa}
\begin{eqnarray}
\text{Model I}: \ \ \ 
 Q = 
\alpha H \rho_m,
\label{model1}
\end{eqnarray}
where $\alpha$
 is a dimensionless parameter. In this scenario the interaction rate is 
proportional to the energy density of DM which is expected to be the case since  
interactions in physics and chemistry are typically stronger if the number 
density of the involved  particles increases. Nevertheless, the interaction 
rate is proportional to the Hubble function too, which although very 
convenient at the calculation stage raises the criticism that the 
DM-DE interaction rate, which is expected to depend on local physics, depends 
on a global feature of the whole Universe, namely the Hubble function 
\cite{Valiviita:2008iv,Shafieloo:2016bpk,Li:2019san}.  Hence, in this work we will additionally consider a 
second interaction function of the local form   
\cite{Valiviita:2008iv}
  \begin{eqnarray}
  \text{Model II}: \ \ \ 
 Q = \Gamma \rho_m,
 \label{model2}
\end{eqnarray} 
where $\Gamma$ is the dimensionfull coupling parameter. Finally, concerning the 
potential choices, we consider two forms 
that have been widely used in the literature,   namely the well known 
exponential potential 
\cite{Copeland:1997et,Amendola:1999er,Elizalde:2004mq,Chen:2008ft,
Basilakos:2011rx,Halder:2024gag} and the hyperbolic potential 
\cite{Paliathanasis:2015gga}, given by 
 \begin{eqnarray}
&&V(\phi)=V_0 e^{-\kappa\lambda\phi}, \label{exp-potential}\\
&&V(\phi)=V_0
\cosh(\kappa\eta\phi), \label{cosh-potential}
\end{eqnarray} 
where $V_0>0$, and $\lambda$, $\eta$    are dimensionless free 
parameters.  The third potential in this series is assumed to be
\cite{Ivanov:2021chn} 
\begin{eqnarray}
&&V 
(\phi) = \frac{1}{8 \beta} \left(1 - e^{- \kappa\mu \phi}\right)^2,  
\label{another-exp-type-potential}
\end{eqnarray} 
with $\mu$, $\beta$ dimensionless parameters.  Let us mention here 
that such type of potential is not purely  phenomenological, since similar 
forms   may be derived   in the context of $R^2$ gravity 
~\cite{Ivanov:2021chn}. 
Finally,  
we will 
also consider the following potential~\cite{Sebastiani:2013eqa}
\begin{eqnarray}
&&V(\phi)=c_0+c_1 e^{\sqrt{\frac{2\kappa^2}{3}}\phi}+c_2 
e^{2\sqrt{\frac{2\kappa^2}{3}}\phi}, \label{gen-exp-potential}
\end{eqnarray} 
with $c_0$,$c_1$,$c_2$, being the non-negative model parameters. Similarly to the 
previous potential, this potential choice can be inspired from $R + R^2 
+\Lambda$ gravity \cite{Sebastiani:2013eqa}, and hence it may 
lead to interesting cosmological implications. 
We mention  that  although the above potentials 
have 
been inspired by higher-order theories of  gravity, such as the $R^2$ one, in 
the present analysis we handle them as scalar-field potentials in the 
framework of General Relativity.  
Note that the last two potentials 
have not been studied in detail in the framework of phantom cosmology.

In summary, the background evolution of the scenario at hand is determined by 
the two Friedmann equations (\ref{friedmann-1}),(\ref{friedmann-2}), together 
with the conservation equations (\ref{cons-CDM}),(\ref{cons-DE}) and the above 
choices for $Q$ and $V (\phi)$. As usual, one  can introduce the dark matter and dark 
energy density parameters as
\begin{eqnarray}
	\Omega_\phi
&\equiv&\frac{\kappa^2\rho_\phi}{3H^2}  \label{eq:Omp_pertds}\\
	\Omega_m &\equiv& \frac{\kappa^2\rho_m}{3 H^2}\,, \label{eq:Omm_pertds}
\end{eqnarray}
 as well as the total, effective, equation-of-state 
parameter as
\begin{eqnarray}
	w_{tot} &\equiv 
&\frac{p_\phi+p_m}{\rho_\phi+\rho_m}=\frac{-\frac{1}{2}\dot{\phi}^2-V(\phi)+w_m 
\rho_m}{-\frac{1}{2}\dot{\phi}^2+V(\phi)+\rho_m}\,.
\end{eqnarray}

\section{Dynamical system analysis}
\label{sec-dyn-systems}

In this section we will perform a detailed phase-space and stability analysis 
 of interacting phantom cosmology. This approach allows one to obtain 
information on the  global dynamics  and the behavior of a cosmological 
scenario,
independently of the initial conditions and the specific universe evolution  
\cite{Copeland:1997et}. As usual, one first transforms the equations  into an 
autonomous dynamical system, and  extracts its critical points. Then, for each 
critical point   one  determines its stability by examining the sign  
of the 
eigenvalues of the perturbation matrix. The dynamical system analysis of 
interacting phantom cosmology  has been performed in various articles 
\cite{Guo:2004xx,Chen:2008ft,Leon:2009dt,Shahalam:2017fqt, 
Chakraborty:2020yfe,Halder:2024gag}, however, it was restricted in specific 
potential choices. In this work we will perform it considering new potential 
forms and interaction models.  In the following we focus on the case  $w_m  = 
0$, namely DM is  cold, and we examine the various 
potentials and interaction forms separately.

\begin{table*}[]
\centering
\resizebox{1.0\textwidth}{!}{%
	\begin{tabular}{c c c c c c c c c}\hline\hline
          
		Point & $x$ & $y$ & Existence & $E_1$ & $E_2$ & Stability & 
$\Omega_\phi$ & 
$w_{tot}$\\ \hline
   &&&&&&&&  \\
            &&&&&&&&  \\
		$A_1$ & $-\frac{\lambda}{\sqrt{6}}$ & $\sqrt{1+\frac{\lambda^2}{6}}$ & 
$\forall$ $\alpha$, $\lambda$ & $-(3+\frac{\lambda^2}{2})$
& $-(\lambda^2+\alpha+3)$ & stable node for $-\alpha<3+\lambda^2$ & $1$ & 
$-1-\frac{\lambda^2}{3}$\\
   &&&&&&&&  \\
            &&&&&&&&  \\
		 &       &      &  for $-3<\alpha<0$   &  &  &   &   &        \\
  
        $A_2$ &  $\frac{3+\alpha}{\sqrt{6}\lambda}\ \ $  &  
$\sqrt{-\frac{\alpha}{3}-\frac{(3+\alpha)^2}{6\lambda^2}}$   &  
 $\lambda^2\geq\frac{(3+\alpha)^2}{-\alpha}$   & 
$\frac{R+\sqrt{S}}{4(3+\alpha)\lambda^2}$ 
 & $\frac{R-\sqrt{S}}{4(3+\alpha)\lambda^2}$  & stable node for $\alpha<-3$    
& 
 
$-\frac{\alpha\lambda^2+(\alpha+3)^2}{3\lambda^2}$ & $\frac{\alpha}{3}$   
   \\

            &     &     & and for $\alpha<-3$  &    &     &   with 
$-(\alpha+3)>\lambda^2\geq\frac{(\alpha+3)^2}{-\alpha}$  &       &      \\
            
             &    &      &   
$-(\alpha+3)\geq\lambda^2\geq\frac{(\alpha+3)^2}{-\alpha}$   &
   &   & (see Fig.  \ref{fig1:model-I})
  &      &      \\
   &&&&&&&&  \\
            &&&&&&&&  \\
\hline\hline
	\end{tabular}%
 }
	\caption{The critical points for the case of  scalar-field potential 
$V(\phi)=V_0 e^{-\kappa\lambda\phi}$ and interaction rate  $Q=\alpha H \rho_m$, 
alongside their existence, eigenvalues at these points and stability 
conditions. 
Additionally, we present 
the corresponding values of the  dark-energy density parameter
$\Omega_\phi$ and the total equation-of-state parameter $w_{tot}$. We 
have defined  
$S=-1944\lambda^2-3240\alpha\lambda^2-2160\alpha^2\lambda^2-720\alpha^3\lambda^2
-120\alpha^4\lambda^2-8\alpha^5\lambda^2-567\lambda^4-1404\alpha\lambda^4-882
\alpha^2\lambda^4-204\alpha^3\lambda^4-15\alpha^4\lambda^4-180\alpha\lambda^6-72
\alpha^2\lambda^6-4\alpha^3\lambda^6+4\alpha^2\lambda^8$ and 
$R=-9\lambda^2+6\alpha\lambda^2+3\alpha^2\lambda^2+2\alpha\lambda^4$.   }
	\label{first-table-potI-modelI}
\end{table*}

\subsection{Potential $V(\phi)=V_0 e^{-\kappa\lambda\phi}$ }

 In this subsection we focus on the  analysis of the case of the exponential 
potential (\ref{exp-potential}) and the two interaction models, one 
corresponding to the global interaction  (\ref{model1}), and one 
corresponding to the local interaction (\ref{model2}). 

\subsubsection{Model I: $Q=\alpha H \rho_m$}

In order to transform the cosmological equations into an autonomous form 
 we introduce the dimensionless variables  
\begin{eqnarray}\label{dim-variables}
   x = \frac{\kappa\dot{\phi}}{\sqrt{6}H},~~  y = 
\frac{\kappa\sqrt{V}}{\sqrt{3}H}.
\end{eqnarray}
Using these variables we can express the density parameters as
\begin{eqnarray}
  \Omega_\phi=-x^2+y^2, ~~~~~ \Omega_m=1+x^2-y^2, 
\end{eqnarray}
while for the total equation-of-state parameter we have
\begin{eqnarray}
        w_{tot}=-x^2-y^2.
\end{eqnarray}
Finally, it proves convenient to introduce the
  deceleration parameter $q\equiv -1-\frac{\dot{H}}{H^2}$, which in the case 
of dust matter becomes simply
$q=\frac{3 w_{tot}+1}{2}$, and thus in terms of the dimensionless variables:
\begin{eqnarray}
 q=\frac{1}{2}\left[1-3(x^2+y^2)\right].
\end{eqnarray}

In terms of  the dimensionless variables (\ref{dim-variables}), the autonomous 
system takes the form:
\begin{eqnarray}
x' &=& -3x-\frac{\sqrt{6}}{2}\lambda y^2+\frac{3}{2}x\left(1-x^2-y^2\right)  
\nonumber\\ && - 
\frac{\alpha}{2x}\left(1+x^2-y^2\right),\label{aut-sys-potI-modelI-1}\\
y' &=& -\frac{\sqrt{6}}{2}\lambda x y+\frac{3}{2}y\left(1-x^2-y^2\right),  
\label{aut-sys-potI-modelI-2}
\end{eqnarray}
where primes denote derivatives with respect to       $N\equiv\ln a$. 
The above system   is invariant under 
the transformation $y\longrightarrow-y$, therefore the phase space is 
defined as
\begin{eqnarray}\label{phy-reg-modelI}
     \mathbb{D}=\left\{ (x,y)\in \mathbb{R}^2: 0\leq-x^2+y^2\leq1, 
y\geq0\right\}.
\end{eqnarray}

Let us make here the following comment. As the
denominator in the evolution equation (\ref{aut-sys-potI-modelI-1}) contains $x$ explicitly, the system exhibits a singularity along
the line $x=0$, which implies that the system becomes ill defined along the line $x=0$. Now, as  $x=0$ marks a singular line, we can partition 
the domain $\mathbb{D}$ into two regions: one for $x>0$ and the other for $x<0$, denoted as 
$\mathbb{D}_+=\mathbb{D}\setminus\{x\leq 0\}$ and 
$\mathbb{D}_-=\mathbb{D}\setminus\{x\geq 0\}$, respectively. After applying a 
re-parametrization, as outlined in Appendix \ref{appendix-1}, one can use  
blow-up techniques \cite{dumortier2006qualitative} to investigate the asymptotic dynamics of the trajectories near the singular line $x=0$ of the system (\ref{aut-sys-potI-modelI-1})-(\ref{aut-sys-potI-modelI-2}) in the regions $\mathbb{D}_+$ and $\mathbb{D}_-$. Subsequently, the nature of the trajectories near the singular line $x=0$ of the system (\ref{aut-sys-potI-modelI-1})-(\ref{aut-sys-potI-modelI-2}) can be characterized by blow-down. This itself is a subject for further investigation. Therefore, in the present work we exclude $x = 0$ from our discussion.

In order to  extract the critical points,  we  impose
$x'=0$ and $y'=0$. The critical points corresponding to an expanding 
Universe (namely having $y>0$), their existence conditions, their eigenvalues 
and their 
stability conditions  are given in 
Table \ref{first-table-potI-modelI}, alongside the corresponding values of the 
 physically interesting 
quantities 
$\Omega_\phi$ and $w_{\rm tot}$. 
As we observe, for this scenario we have the following critical points.

\begin{figure}
	\includegraphics[width=0.4\textwidth]{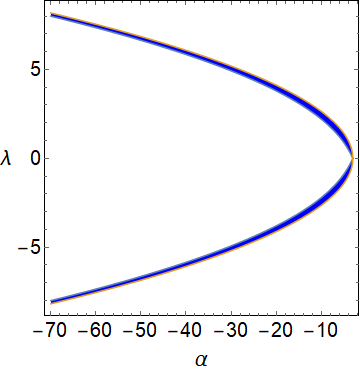}
	\caption{{\it{The stability region of critical point $A_2$,
for the case of scalar-field potential 
$V(\phi)=V_0 e^{-\kappa\lambda\phi}$ and interaction rate $Q=\alpha H \rho_m$. } }} 
	\label{fig1:model-I}
\end{figure}

\begin{itemize}
    \item  The critical point $A_1$ exists for all values of the parameters 
$\alpha$ and $\lambda$. It represents the dark-energy dominated 
($\Omega_\phi=1$) solution with 
total equation of state $-1-\frac{\lambda^2}{3}$, and thus it corresponds   
to a phantom accelerating universe. It is a stable node when the condition 
$-\alpha<3+\lambda^2$ is satisfied, and as a  late-time attractor  this point 
can describe the late-time 
universe. 

  \item The critical point $A_2$ exists in the parameter region shown 
in Table \ref{first-table-potI-modelI}. Since the eigenvalues of the 
corresponding linearized matrix  are complicated, the stability conditions have 
to be determined  numerically, and are presented in 
  Fig.  \ref{fig1:model-I}. 
When  $\alpha<-1$, it corresponds to an accelerating 
universe. Moreover, at   $A_2$  we have
$\Omega_m/\Omega_{\phi} \neq 0$, and thus this point represents late-time 
accelerating scaling solution.
Hence, in this case, the coincidence problem can be alleviated, as reported 
earlier in \cite{Chen:2008ft}.
Additionally, we  mention that the critical points $A_2$ can represent 
purely matter dominated solutions (namely having $\Omega_m =1$) for the 
parameters $\lambda$ and $\alpha$ satisfying $\lambda^2=-(\alpha+3)^2/\alpha$.  
 Thus, depending on the parameter values, point $A_2$ can correspond to the above
possibilities. As we shall see below, for different potential  choices such 
late-time stable accelerating scaling solutions can be obtained in a larger 
parameter region compared to this case.  
    
\end{itemize}

We close this subsection by mentioning that due to the 
non-compactness of the phase space in the $x-y$ direction discussed above, we 
should examine whether there are 
critical points at infinity. Hence, to complete the analysis, we compactify the phase space $\mathbb{D}$ by employing the Poincar\'{e} compactification 
technique 
\cite{perko2013differential,dumortier2006qualitative,meiss2007differential, 
Bahamonde:2017ize} and study the qualitative behaviour at the critical points at infinity. The details are provided in Appendix 
\ref{appendix-1}.  Our analysis reveals that the system possesses two critical
points at infinity, namely
$A^{\pm\infty}\left(\pm\frac{1}{\sqrt{2}},\frac{1}{\sqrt{2}},0\right)$, which
exhibit unstable behavior, and thus they cannot attract the Universe at late 
times. Additionally,  we can see that at these critical points,
$q\longrightarrow-\infty$. However, note that this is not considered as a Big 
Rip, since it is obtained at infinite rather than at finite time (see the 
classifications of singularities in a phantom Universe, in 
\cite{Nojiri:2005sx}). 

\begin{table*}[]
\centering
 {%
{%
\begin{tabular}{c c c c c c c c c c c }\hline\hline
          
		Point & $x$ & $y$ & $z$  & Existence & $E_1$ & $E_2$ & $E_3$ 
& 
Stability & $\Omega_\phi$ & $w_{tot}$\\ \hline
   &&&&&&&&&&  \\
            &&&&&&&&&&  \\
		$B_1$ & $-\frac{\lambda}{\sqrt{6}}$ &  $\sqrt{1+\frac{\lambda^2}{6}}$ & 
0 & $\forall$ $\gamma$, $\lambda$ & $-(3+\lambda^2)$ & 
$-(\frac{\lambda^2}{2}+3)$ & $-\frac{\lambda^2}{2}$  & 
always stable node  & 1 & $-1-\frac{\lambda^2}{3}$\\
   &&&&&&&&&  \\
            &&&&&&&&&  \\
         $B_2$ & $-\frac{\lambda}{\sqrt{6}}$ &  $\sqrt{1+\frac{\lambda^2}{6}}$ 
& 
1  & $\forall$ $\gamma$, $\lambda$ 
    & $-sgn(\gamma)\infty$ & $-(\frac{\lambda^2}{2}+3)$ & $\frac{\lambda^2}{2}$ 
& always saddle  & 1 & $-1-\frac{\lambda^2}{3}$\\
          &&&&&&&&&  \\
            &&&&&&&&&  \\
		 \hline\hline
	\end{tabular}}
}
	\caption{
	The critical points for the case of  scalar-field potential 
$V(\phi)=V_0 e^{-\kappa\lambda\phi}$ and interaction rate $Q=\Gamma \rho_m$, 
alongside their eigenvalues and their existence and stability conditions. 
Additionally, we present 
the corresponding values of the  dark-energy density parameter
$\Omega_\phi$ and the total equation-of-state parameter $w_{tot}$.    }
	\label{first-table-potI-modelII}
\end{table*}

\subsubsection{Model II: $Q=\Gamma\rho_m$}

In this case, apart from the dimensionless variables
 (\ref{dim-variables}) we need one more, namely  
\begin{eqnarray}\label{dim-variable-z}
  z=\frac{H_0}{H+H_0}, 
\end{eqnarray}
 with $H_0$ the value of the Hubble function at present. Hence, the 
autonomous system becomes: 
\begin{eqnarray}
x'&=&-3x-\frac{\sqrt{6}}{2}\lambda y^2+\frac{3}{2}x\left(1-x^2-y^2\right)  
\nonumber \\ && -\frac{\gamma z 
\left(1+x^2-y^2\right)}{2x(1-z)},\label{aut-sys-potI-modelII-1}\\
y' &=& -\frac{\sqrt{6}}{2}\lambda x y+\frac{3}{2}y\left(1-x^2- y^2\right),\\ 
\label{aut-sys-potII-modelII-2}
z' &=& \frac{3}{2}z(1-z)\left(1-x^2-y^2\right),\label{aut-sys-potI-modelII-3}
\end{eqnarray}
where $\gamma=\Gamma/H_0$ is the dimensionless coupling parameter. It is clear 
that the autonomous system 
(\ref{aut-sys-potI-modelII-1})-(\ref{aut-sys-potI-modelII-3}) is invariant 
under 
the transformation $y\longrightarrow-y$, and therefore the physical region is 
defined as {\small{
\begin{equation}
 \label{phy-reg-modelII}
    \mathbb{D}=\bigg\{ (x,y,z)\in \mathbb{R}^3: 0\leq-x^2+y^2\leq1, 
  y\geq 
0, 0\leq z \leq1 \bigg\}.
\end{equation}}}

Similar to the previous subsection, here we also observe that the explicit dependence of the denominator in equation (\ref{aut-sys-potI-modelII-1})  on $x$  leads to a singularity at $x=0$ plane, rendering the above system ill-defined at this 
plane. This issue can be addressed in an asymptotic way by dividing  
the entire phase space $\mathbb{D}$ into two separate domains, namely, 
$\mathbb{D}_+=\mathbb{D}\setminus\{x\leq 0\}$ and 
$\mathbb{D}_-=\mathbb{D}\setminus\{x\geq 0\}$,
and by using re-parametrization and suitable blow-up transformations.
Similarly, equation (\ref{aut-sys-potI-modelII-1}) features $(1-z)$ in the 
denominator, resulting in a singularity at the $z=1$ plane, which   makes the 
system ill-defined at this location too. To resolve this,  the system 
can be regularized by introducing the factor $(1-z)$. However, the analysis of 
the regularized dynamical system and the technique utilized in this article are 
equivalent (see  
\cite{Boehmer:2008av,Khyllep:2021wjd,Bahamonde:2017ize,Halder:2024gag, 
Halder:2024uao}), and therefore, the regularized system does not add any new 
physical implications in this context. Note that the same methodology can be 
applied to all autonomous systems in the following where the dynamical 
variables are in  the denominators.

The critical points, their existence conditions, their eigenvalues  and their 
stability conditions  are given in    Table \ref{first-table-potI-modelII}, 
alongside the values of  $\Omega_\phi$ and $w_{tot}$.
As we observe, for this scenario we have the following critical points.

\begin{itemize}
    \item  The critical point $B_1$ exists always and corresponds to 
dark-energy 
dominated solutions. Since the total equation-of-state 
parameter is always less than $-1$, it corresponds to a phantom   
universe. Since, all the eigenvalues evaluated at $B_1$ are negative, it is 
always stable node.  Thus, this point depicts a  dark-energy dominated 
accelerated solution. However, note that it
corresponds to  $H\longrightarrow\infty$, as expected for a phantom, 
super-accelerating (i.e. with $\dot{H}>0$) universe.
    
  \item As before, point $B_2$ corresponds to a dark-energy dominated 
solution, and exists always. It is  always saddle and it has
 $H \longrightarrow 0$.  Note that since it differs from point $B_1$ only 
in the $z$ value, both $B_1$ and $B_2$ have the same $\Omega_\phi$ and 
$w_{tot}$.

\end{itemize} 

Finally, since  the phase space is not compact in the $x-y$ 
directions, we expect to have critical points at infinity. 
In order to provide a complete analysis we apply the Poincar\'{e} 
compactification technique to examine the stability at these critical points. 
This analysis shows that the system has two points at infinity, namely
$B^{\pm\infty}\left(\pm\frac{1}{\sqrt{2}},\frac{1}{\sqrt{2}},0,0\right)$, both 
of which display unstable behavior, and thus they cannot attract the Universe at 
late 
times. The details are presented in Appendix 
\ref{appendix-2}. At these points one can observe that 
$q\longrightarrow-\infty$.

Finally, we mention that this local interaction model  was investigated in 
\cite{Chen:2008ft}, considering the exponential potential, and the absence of 
  accelerating  scaling attractors was also reported.

\begin{table*}[]
\centering
\resizebox{1.03\textwidth}{!}{%
\begin{tabular}{c c c c c c c c c c c }\hline\hline
          
		Point & $x$ & $y$ & $\lambda$  & Existence & $E_1$ & $E_2$ & $E_3$ & 
Stability & 
$\Omega_\phi$ & $w_{tot}$\\ \hline
   &&&&&&&&&&  \\
            &&&&&&&&&&  \\
		$C_1$ & $-\frac{\eta}{\sqrt{6}}$ & $\sqrt{1+\frac{\eta^2}{6}}$ & $\eta$ 
& $\forall$ $\alpha$, $\eta$ & $-(3+\alpha+\eta^2)$  & $-(3+\frac{\eta^2}{2})$  
& $-2\eta^2$  & stable node for $-\alpha<3+\eta^2$ & 1 & $-1-\frac{\eta^2}{3}$\\
   &&&&&&&&&&  \\
            &&&&&&&&&&  \\
		 &       &      &     & for { $-3<\alpha<0$} &  &  &  &   &   &        
\\
  
        $C_2$ &  $\frac{3+\alpha}{\sqrt{6}\eta}$  ~&~  
$\sqrt{-\frac{\alpha}{3}-\frac{(3+\alpha)^2}{6\eta^2}}$ ~&~ $\eta$     ~&~  
$\eta^2\geq\frac{(3+\alpha)^2}{-\alpha}$ & 
$\frac{P+\sqrt{S}}{4(3+\alpha)\eta^2}$ & $\frac{P-\sqrt{S}}{4(3+\alpha)\eta^2}$ 
 
& $2(3+\alpha)$ & stable node for 
$\alpha<-3$    &   $-\frac{\alpha\eta^2+(\alpha+3)^2}{3\eta^2}$ & 
$\frac{\alpha}{3}$      \\

            &     &     &      &  and for { $\alpha<-3$}   &  &  &  &
$-(\alpha+3)>\eta^2\geq\frac{(\alpha+3)^2}{-\alpha}$  &       &      \\
            
             &    &      &      &   
$-(\alpha+3)\geq\eta^2\geq\frac{(\alpha+3)^2}{-\alpha}$   &  &  &  &  (see Fig. 
\ref{fig1:model-II})   &   
 
  &      \\
   &&&&&&&&&&  \\
            &&&&&&&&&&  \\
		$C_3$ & $\frac{\eta}{\sqrt{6}}$ & $\sqrt{1+\frac{\eta^2}{6}}$ & $-\eta$ 
& $\forall$ $\alpha$, $\eta$ & $-(3+\alpha+\eta^2)$ & $-(3+\frac{\eta^2}{2})$ & 
$-2\eta^2$ & stable node for $-\alpha<3+\eta^2$ & 1 & $-1-\frac{\eta^2}{3}$\\
   &&&&&&&&&&  \\
            &&&&&&&&&&  \\
		 &       &      &     & for { $-3<\alpha<0$} &   &  &  &   &   &        
\\
  
        $C_4$ &  $-\frac{3+\alpha}{\sqrt{6}\eta}$  &  
$\sqrt{-\frac{\alpha}{3}-\frac{(3+\alpha)^2}{6\eta^2}}$ & $-\eta$     &  
$\eta^2\geq\frac{(3+\alpha)^2}{-\alpha}$ & 
$\frac{P+\sqrt{S}}{4(3+\alpha)\eta^2}$ & $\frac{P-\sqrt{S}}{4(3+\alpha)\eta^2}$ 
& $2(3+\alpha)$  & stable node for 
$\alpha<-3$  &  $-\frac{\alpha\eta^2+(\alpha+3)^2}{3\eta^2}$ & 
$\frac{\alpha}{3}$      \\

            &     &     &      &  and for { $\alpha<-3$}   &  &  &  & 
$-(\alpha+3)>\eta^2\geq\frac{(\alpha+3)^2}{-\alpha}$  &       &      \\
            
             &    &      &     &   
$-(\alpha+3)\geq\eta^2\geq\frac{(\alpha+3)^2}{-\alpha}$ &  &  &   &  (see Fig. 
\ref{fig1:model-II})   &   
 
  &      \\
   &&&&&&&&&&  \\
            &&&&&&&&&&  \\
     \hline\hline
	\end{tabular}%
 }
	\caption{
		The critical points for the case of  
 $V(\phi)=V_0\cosh{(\kappa\eta\phi)}$ and interaction rate $Q=\alpha 
H \rho_m$,
alongside their existence, their eigenvalues and stability conditions. 
Additionally, we present 
the corresponding values of the  dark-energy density parameter
$\Omega_\phi$ and the total equation-of-state parameter $w_{tot}$. We 
have defined   
$S=-1944\eta^2-3240\alpha\eta^2-2160\alpha^2\eta^2-720\alpha^3\eta^2-120\alpha^4
\eta^
2-8\alpha^5\eta^2-567\eta^4-1404\alpha\eta^4-882\alpha^2\eta^4-204\alpha^3\eta^4
-15
\alpha^4\eta^4-180\alpha\eta^6-72\alpha^2\eta^6-4\alpha^3\eta^6+4\alpha^2\eta^8$
and 
  $P=-9\eta^2+6\alpha\eta^2+3\alpha^2\eta^2+2\alpha\eta^4$. 
  }
	\label{first-table-potII-modelI}
\end{table*}

\subsection{Potential: $V(\phi)=V_0\cosh{(\kappa\eta\phi)}$ }

  In this subsection we focus on the phase space analysis of the autonomous 
system for the hyperbolic potential (\ref{cosh-potential}) considering  the 
global interaction rate  (\ref{model1}) and the local interaction rate 
 (\ref{model2}).   

\subsubsection{Model I: $Q=\alpha H \rho_m$}

In order to transform the cosmological equations into an autonomous form,
  apart from the dimensionless variables
 (\ref{dim-variables}) we need one more, namely   
\begin{eqnarray}\label{dim-variable-lamda}
  \lambda=-\frac{V_{\phi}}{\kappa V},
\end{eqnarray}
 where a $\phi$-subscript denotes derivative with respect to $\phi$.
  If $\lambda (\phi)$ is invertible, 
  and one can extract the form 
$\phi=\phi(\lambda)$, then   the extraction of an autonomous system     can be 
obtained by simply calculating the parameter $\Xi=\frac{V 
V_{\phi\phi}}{{V_{\phi}}^2}$. Using the potential 
$V(\phi)=V_0\cosh{(\kappa\eta\phi)}$, 
we find $\lambda=-\eta\tanh{(\kappa\eta\phi)}$ and 
$\Xi=\frac{\eta^2}{\lambda^2}$ .  Therefore, the range of $\lambda$ is 
$[-|\eta|,|\eta|]$.  Finally, the autonomous system  takes the form:
\begin{eqnarray}
   x' &=& -3x- \frac{\sqrt{6}}{2}\lambda y^2+\frac{3}{2}x\left(1-x^2-y^2\right) 
\nonumber \\ &&-  
\frac{\alpha}{2x}\left(1+x^2-y^2\right),\label{aut-sys-potII-modelI-1}\\
    y' &=& -\frac{\sqrt{6}}{2}\lambda x 
y+\frac{3}{2}y\left(1-x^2-y^2\right),\label{aut-sys-potII-modelI-2}\\
    \lambda' &=&  
\sqrt{6}\left(\lambda^2-\eta^2\right)x.\label{aut-sys-potII-modelI-3}
\end{eqnarray}
Since the above 
 autonomous system  is invariant under 
the transformation $y\longrightarrow-y$,   the phase space domain 
$\mathbb{D}$ is defined as $\mathbb{D}=\left\{ (x,y,\lambda)\in \mathbb{R}^3: 
0\leq-x^2+y^2\leq1, y\geq 0, -|\eta| \leq\lambda\leq |\eta| \right\}$.
The presence of $x$ in the denominator of   equation 
(\ref{aut-sys-potII-modelI-1}) introduces a singularity at $x=0$ plane, causing 
the system to become ill-defined at this plane, and this can be resolved in a similar fashion as discussed in the previous subsections.

 The critical points, their existence conditions, their eigenvalues and their 
stability conditions are given in 
Table \ref{first-table-potII-modelI}, alongside the corresponding values of  
$\Omega_\phi$ and $w_{tot}$.  
For this scenario we have the following critical points.

\begin{figure}[!]
	\includegraphics[width=0.4\textwidth]{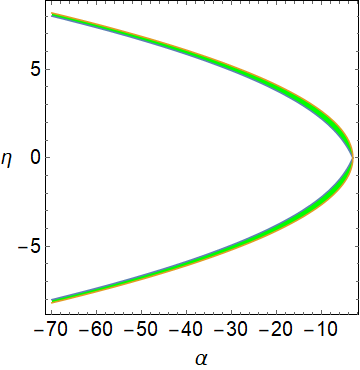}
	\caption{{\it{The stability region of the critical points $C_2$ and $C_4$,
for the case of scalar-field potential 
 $V(\phi)=V_0\cosh{(\kappa\eta\phi)}$ and interaction rate $Q=\alpha H \rho_m$. 
} }} 
	\label{fig1:model-II}
\end{figure}

\begin{itemize}
    \item   The critical point $C_1$ exists for all $\alpha$ and $\eta$ and it 
corresponds to an accelerating universe, and it is stable in the region 
$-\alpha<3+\eta^2$.  Moreover, at this 
point we have  $\Omega_\phi=1$, which implies  complete dark energy 
domination. Therefore, it describes the late-time   dark-energy dominated 
accelerated universe. 

\item The critical point $C_2$ describes an accelerating universe, since  
in their stability region, i.e. for $\alpha<-1$, we have $w_{tot}<-1/3$.   
The stability region is graphically shown in Fig. 
\ref{fig1:model-II}. As we observe, in this case we find almost 
similar stability region as in Fig. \ref{fig1:model-I}   for  the exponential 
potential, which was  expected since the present hyperbolic cosinus potential 
is related to the  exponential potential.

Again, we notice that at $C_2$  the 
ratio of $\Omega_m$ and $\Omega_\phi$ is non zero (namely $\Omega_m/\Omega_\phi 
\neq 0$),   and thus this   point represents a late-time accelerating scaling 
solution, which is stable   in the aforementioned specific parameter region. Thus, for this case, the coincidence 
problem can be alleviated. Finally, one can further notice that $C_2$ 
can represent completely matter dominated solutions    for a 
specific parameter space satisfying $\eta^2=-(\alpha+3)^2/\alpha$.

\item The   critical point $C_3$ is stable for
$-\alpha<3+\eta^2$, and the values of $\Omega_{\phi}$ and 
$w_{\rm tot}$ are $1$ and $-1-\frac{\eta^2}{3}<-1$ respectively,   
therefore this point represents dark-energy dominated accelerated solutions. 
Hence, point $C_3$, similar to $C_1$, is a candidate 
for the late-time universe.

\item The critical point $C_4$ has 
$w_{\rm tot}=\frac{\alpha}{3}<-\frac{1}{3}$ for $\alpha<-1$, and thus it 
corresponds to accelerated solutions.   The stability region is
graphically shown in Fig. \ref{fig1:model-II}, which is the same with point 
$C_2$ since they have the same eigenvalues.
At this critical 
point $\Omega_m/\Omega_{\phi} \neq 0$ and therefore it corresponds to
late-time accelerating stable solutions that can alleviate the 
coincidence problem. Finally, completely matter-dominated 
solutions  are obtained for
$\eta^2=-(\alpha+3)^2/\alpha$.  

\end{itemize}

Finally, investigating the qualitative behaviour at the critical points at infinity   (see   Appendix \ref{appendix-3}) reveals that the phase space domain $\mathbb{D}$ contains four such critical points, namely   
$C_1^{\pm\infty}\left(\pm\frac{1}{\sqrt{2}},\frac{1}{\sqrt{2}},0,0\right)$ and 
$C_2^{\pm\infty}\left(\pm\sqrt{\frac{3}{8}},\sqrt{\frac{3}{8}},\mp\frac{1}{2}, 
0\right)$. The first two points are unstable, while the remaining two 
exhibit saddle behavior. At these points, we deduce that 
$q\longrightarrow-\infty$.

\subsubsection{Model II: $Q=\Gamma \rho_m$}

\begin{table*}[]
\centering {%
\begin{tabular}{c c c c c c c c c c c c c  }\hline\hline
          
		Point & $x$ & $y$ & $\lambda$ & $z$ & Existence & $E_1$ & $E_2$ & 
$E_3$ & $E_4$ & Stability & $\Omega_\phi$ & $w_{tot}$\\ \hline
   &&&&&&&&&&&&  \\
            &&&&&&&&&&&&  \\
		$D_1$ & $-\frac{\eta}{\sqrt{6}}$ ~&~  $\sqrt{1+\frac{\eta^2}{6}}$ ~&~ 
$\eta$ ~&~ 
0 ~&~ $\forall$ $\gamma$, $\eta$ & $-(3+\eta^2)$ & $-(\frac{\eta^2}{2}+3)$ & 
$-2\eta^2$ & $-\frac{\eta^2}{2}$ &  stable node  & 1 & 
$-1-\frac{\eta^2}{3}$\\
   &&&&&&&&&&  \\
            &&&&&&&&&&  \\
         $D_2$ & $-\frac{\eta}{\sqrt{6}}$ ~&~  $\sqrt{1+\frac{\eta^2}{6}}$ ~&~ 
$\eta$ 
~&~ 1 ~&~ $\forall$ $\gamma$, $\eta$ 
    & $-sgn(\gamma)\infty$ & $-(\frac{\eta^2}{2}+3)$ & $-2\eta^2$ & 
$\frac{\eta^2}{2}$ &  always saddle  & 1 & 
$-1-\frac{\eta^2}{3}$\\
          &&&&&&&&&&&&  \\
            &&&&&&&&&&&&  \\
         
		$D_3$ & $\frac{\eta}{\sqrt{6}}$ ~&~ $\sqrt{1+\frac{\eta^2}{6}}$ ~&~ 
$-\eta$ ~&~ 
0 ~&~ $\forall$ $\gamma$, $\eta$ & $-(3+\eta^2)$ & $-(\frac{\eta^2}{2}+3)$ & 
$-2\eta^2$ & $-\frac{\eta^2}{2}$ &  stable node  & 1 & 
$-1-\frac{\eta^2}{3}$\\
   &&&&&&&&&&  \\
            &&&&&&&&&&  \\
         $D_4$ & $\frac{\eta}{\sqrt{6}}$ ~&~  $\sqrt{1+\frac{\eta^2}{6}}$ ~&~ 
$-\eta$ 
~&~  1 ~&~ $\forall$ $\gamma$, $\eta$ 
    & $-sgn(\gamma)\infty$ & $-(\frac{\eta^2}{2}+3)$ & $-2\eta^2$ & 
$\frac{\eta^2}{2}$ & always saddle  & 1 & 
$-1-\frac{\eta^2}{3}$\\
          &&&&&&&&&&&&  \\
            &&&&&&&&&&&&  \\   
		 \hline\hline
	\end{tabular}}
	\caption{
			The critical points for the case of  
 $V(\phi)=V_0\cosh{(\kappa\eta\phi)}$ and interaction rate $Q=\Gamma 
\rho_m$, 
alongside their eigenvalues and their existence and stability conditions. 
Additionally, we present 
the corresponding values of the  dark-energy density parameter
$\Omega_\phi$ and the total equation-of-state parameter $w_{tot}$.   
   }
	\label{first-table-potII-modelII}
\end{table*}

Using the dimensionless variables $x$, $y$  defined in  
(\ref{dim-variables}) alongside the dimensionless variables $z$ and $\lambda$  
of (\ref{dim-variable-z}) and (\ref{dim-variable-lamda}), we 
obtain the autonomous system as
\begin{eqnarray}
x' &=& -3x-\frac{\sqrt{6}}{2}\lambda  y^2+\frac{3}{2}x\left(1-x^2-y^2\right) 
\nonumber\\ && -\frac{\gamma z 
\left(1+x^2-y^2\right)}{2x(1-z)},\label{aut-sys-potII-modelII-1}\\ 
y' &=& -\frac{\sqrt{6}}{2}\lambda x y+\frac{3}{2}y\left(1-x^2- y^2\right), 
\label{aut-sys-potII-modelII-2}\\
 \lambda' &=& 
\sqrt{6}\left(\lambda^2-\eta^2\right)x,\label{aut-sys-potII-modelII-3}\\ 
z' &=& 
\frac{3}{2}z(1-z)\left(1-x^2-y^2\right),\label{aut-sys-potII-modelII-4}
\end{eqnarray}
where $\gamma=\Gamma/H_0$ is  the dimensionless coupling parameter. 
The above autonomous system   is invariant under the transformation 
$y\longrightarrow-y$, and the phase space $\mathbb{D}$ is given by 
\begin{eqnarray}\label{phy-reg-modelII}
    \mathbb{D}=\bigg\{ (x,y,\lambda,z) \in \mathbb{R}^4: 0\leq-x^2+y^2\leq1, 
y\geq 0,\nonumber\\ -|\eta|\leq\lambda\leq |\eta|, 0\leq z \leq1 \bigg\}.\ \ 
\end{eqnarray}
Since $x$ and $(1-z)$ appear in the denominator of the evolution 
equation (\ref{aut-sys-potII-modelII-1}), the system exhibits singularities at 
$x=0$ and $z=1$ surfaces, making it ill-defined along these surfaces.

The critical points, their existence conditions, their 
stability conditions, and the values of $\Omega_\phi$ and $w_{tot}$,  are given 
in Table  \ref{first-table-potII-modelII}. We have the following critical 
points.

\begin{itemize}
\item The critical point $D_1$ exists always and it 
corresponds to 
super-accelerated dark-energy domination, exhibiting $H\longrightarrow\infty$. 
It is always stable and it can describe the Universe at late times.

\item The critical point $D_2$ exists always,  corresponding to dark energy 
dominated accelerated solutions, with    $H\longrightarrow 0$, and it is  
always saddle.    

\item Point $D_3$ exists always, and it corresponds to dark-energy 
dominated accelerated solutions, with $H\longrightarrow\infty$.  Since it is 
stable, it is also a suitable candidate for late-time evolution of the 
universe.

\item  Point $D_4$ exists always, it is saddle, and it corresponds to 
dark-energy 
dominated accelerated solutions with
$H \longrightarrow 0$.   
    
\end{itemize}

Finally, with the help of the the Poincar\'{e} compactification it follows that the system encounters two critical points at infinity, namely
$D^{+\infty}\Bigl\{\left(y_1,y_2,y_3,y_4,y_5\right)\in\mathbb{S}^4:y_5=0,y_4=0,
y_1^2=y_2^2, y_1>0,y_2>0\Bigr\}$, and 
$D^{-\infty}\Bigl\{\left(y_1,y_2,y_3,y_4,y_5\right)\in\mathbb{S}^4:y_5=0,y_4=0,
y_1^2=y_2^2, y_1<0,y_2>0\Bigr\}$, where $q\longrightarrow-\infty$. The analysis reveals that these 
critical points exhibit unstable behavior. The details of the analysis are 
presented in Appendix \ref{appendix-4}.  

In summary, the hyperbolic potential with the local interaction rate leads to 
results similar to the exponential potential, since the two potentials are 
related.  
Thus, in this scenario, late-time scaling attractors are not possible.

\subsection{Potential: $V (\phi) = \frac{1}{8 \beta} \left(1 - e^{- \kappa 
\mu\; 
\phi}\right)^2$}

 In this subsection we focus on the phase space analysis of the autonomous 
system for the potential (\ref{another-exp-type-potential}), considering as 
usual 
both the global interaction rate   (\ref{model1})  and the local interaction 
rate  (\ref{model2}).    

\subsubsection{Model I: $Q=\alpha H \rho_m$}
\begin{table*}[]
\centering
\resizebox{1.0\textwidth}{!}{%
	\begin{tabular}{c c c c c c c c c c c }\hline\hline
          
		Point & $x$ & $y$ & $\lambda$  & Existence & $E_1$ & $E_2$ & $E_3$ & 
Stability & 
$\Omega_\phi$ & $w_{tot}$\\ \hline
   &&&&&&&&&&  \\
            &&&&&&&&&&  \\
		$F_1$ & $-\frac{2\mu}{\sqrt{6}}$ & $\sqrt{1+\frac{2\mu^2}{3}}$ & $2\mu$ 
 
& $\forall$ $\alpha$, $\mu$ & $-(3+\alpha+4\mu^2)$ & $-3-2\mu^2$ & $-2\mu^2$ & 
stable node for $-\alpha<3+4\mu^2$ & 1 & $-1-\frac{4\mu^2}{3}$\\
   &&&&&&&&&&  \\
            &&&&&&&&&&  \\
		 &       &      &    & for { $-3<\alpha<0$} &   &  &  &  &   &        \\
  
        $F_2$ &  $\frac{3+\alpha}{2\sqrt{6}\mu}$  &  
$\sqrt{-\frac{\alpha}{3}-\frac{(3+\alpha)^2}{24\mu^2}}$ & $2\mu$  &  
$\mu^2\geq\frac{(3+\alpha)^2}{-4\alpha}$ & 
$\frac{P+\sqrt{S}}{4(3+\alpha)\mu^2}$ 
& $\frac{P-\sqrt{S}}{4(3+\alpha)\mu^2}$ & $\frac{1}{2}(3+\alpha)$ & stable node 
for 
$\alpha<-3$  & $-\frac{4\alpha\mu^2+(\alpha+3)^2}{12\mu^2}$ & 
$\frac{\alpha}{3}$      \\

            &     &     &     &  and for { $\alpha<-3$}   &  &  &  &
$-\frac{(\alpha+3)}{4}>\mu^2\geq\frac{(\alpha+3)^2}{-4\alpha}$  &       &      
\\
            
             &    &      &     &   
$-\frac{(\alpha+3)}{4}\geq\mu^2\geq\frac{(\alpha+3)^2}{-4\alpha}$   &  &  &  & 
(see Fig. \ref{fig1-model-III})   &      &      \\
   &&&&&&&&&&  \\
            &&&&&&&&&&  \\
		{ $F_3$} & $x_c$ & $\sqrt{1-x_c^2}$ & $0$ 
 & $\forall\mu$, $\alpha=-3$, & $0$  & $-6$  & $-\sqrt{6}\mu x_c$  &  
stable for   $\mu >0  $ and $ x_c >0$     & $1-2x_c^2$ & 
$-1$\\
   &&&& $|x_c|\leq\frac{1}{\sqrt{2}}$ with $x_c\neq 0$ &  &  &  & or for $\mu<0
   $ and $x_c 
<0$  &&  \\
            &&&&&&&&&&  \\
		 \hline\hline
	\end{tabular}%
 }
	\caption{The critical points for the case of  
 $V (\phi) = \frac{1}{8 \beta} \left(1 - e^{- \kappa \mu\; 
\phi}\right)^2$ and interaction rate $Q=\alpha H \rho_m$,
alongside their existence, eigenvalues and stability conditions. Furthermore, 
we 
present 
the corresponding values of the  dark-energy density parameter
$\Omega_\phi$ and the total equation-of-state parameter $w_{tot}$. We 
have defined    
$S=-486\mu^2-810\alpha\mu^2-540\alpha^2\mu^2-180\alpha^3\mu^2-30\alpha^4\mu^2-2
\alpha^5\mu^2-567\mu^4-1404\alpha\mu^4-882\alpha^2\mu^4-204\alpha^3\mu^4-15
\alpha^4\mu^4-720\alpha\mu^6-288\alpha^2\mu^6-16\alpha^3\mu^6+24\alpha^2\mu^8$ 
and 
  $P=-9\mu^2+6\alpha\mu^2+3\alpha^2\mu^2+8\alpha\mu^4$.   }
	\label{first-table-potIII-modelI}
\end{table*}

In this case we derive the autonomous system in terms of the 
dimensionless variables $x$, $y$, $\lambda$,  defined in 
(\ref{dim-variables}) and (\ref{dim-variable-lamda}). As we can see, the 
expressions of the parameters $\lambda=-\frac{V_{\phi}}{\kappa V}$ and 
$\Xi=\frac{V V_{\phi\phi}}{V_{\phi}^2}$ take  the form 
$\lambda=-2\mu\frac{e^{-\kappa\mu\phi}}{1-e^{-\kappa\mu\phi}}$ ($\mu\neq0$) and 
$\Xi=\frac{\mu}{\lambda}+\frac{1}{2}$, respectively. Therefore, the autonomous 
system becomes:
\begin{eqnarray}
   x' &=&-3x-\frac{\sqrt{6}}{2}\lambda y^2+\frac{3}{2}x\left(1-x^2-y^2\right)  
\nonumber\\ && -  
\frac{\alpha}{2x}\left(1+x^2-y^2\right),\label{aut-sys-potIII-modelI-1}\\
    y' &=& -\frac{\sqrt{6}}{2}\lambda x y+\frac{3}{2}y\left(1-x^2-y^2\right),\\ 
\label{aut-sys-potIII-modelI-2}
    \lambda' &=& \frac{\sqrt{6}}{2}\left(\lambda^2-2\mu \lambda \right)x. 
\label{aut-sys-potIII-modelI-3}
\end{eqnarray}
The above  system  is invariant under $y\longrightarrow-y$, and its phase 
space region $\mathbb{D}$ is defined by $\mathbb{D}=\left\{ (x,y,\lambda)\in 
\mathbb{R}^3: 0\leq-x^2+y^2\leq1, y\geq 0\right\}$.
The explicit presence of $x$ in the denominator of equation 
(\ref{aut-sys-potIII-modelI-1}) leads to a singularity at $x=0$ plane, where the 
system is no longer well-defined.

 The critical points, their existence conditions, their eigenvalues and their 
stability conditions are given in 
Table \ref{first-table-potIII-modelI}, alongside the corresponding values of  
$\Omega_\phi$ and $w_{tot}$.  
We have the following critical 
points.
\begin{figure}[!]
\includegraphics[width=0.4\textwidth]{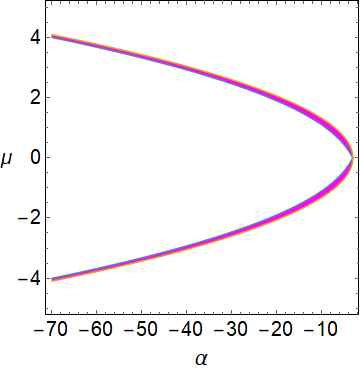}
\caption{{\it{The   stability region of the critical point 
$F_2$,
for the case of scalar-field potential 
$V (\phi) = \frac{1}{8 \beta} \left(1 - e^{- \kappa \mu\; 
\phi}\right)^2$ and interaction rate $Q=\alpha H \rho_m$. 
} }} 
	\label{fig1-model-III}
\end{figure}

\begin{table*}[]
\centering
\resizebox{1.0\textwidth}{!}{%
	\begin{tabular}{c c c c c c c c c c c c c}\hline\hline
          
		Point & $x$ & $y$ & $\lambda$ & $z$ & Existence & $E_1$ & $E_2$ & 
$E_3$ & $E_4$  & Stability & $\Omega_\phi$ & $w_{tot}$\\ \hline
   &&&&&&&&&&&&  \\
            &&&&&&&&&&&&  \\
		$G_1$ & $-\frac{2\mu}{\sqrt{6}}$ &  $\sqrt{1+\frac{2\mu^2}{3}}$ & 
$2\mu$ 
& 0 & $\forall$ $\gamma$, $\mu$ & $-3-4\mu^2$ & $-2\mu^2-3$ & $-2\mu^2$ & 
$-2\mu^2$ &  stable node  & 1 & $-1-\frac{4\mu^2}{3}$\\
   &&&&&&&&&&&&  \\
            &&&&&&&&&&&&  \\
         $G_2$ & $-\frac{2\mu}{\sqrt{6}}$ &  $\sqrt{1+\frac{2\mu^2}{3}}$ & 
$2\mu$ & 1  & $\forall$ $\gamma$, $\mu$ 
    & $-sgn(\gamma)\infty$ & $-2\mu^2-3$ & $-2\mu^2$ & $2\mu^2$  & 
always saddle  & 1 & $-1-\frac{4\mu^2}{3}$\\
          &&&&&&&&&&&&  \\
            &&&&&&&&&&&&  \\
         
		$G_3$ & $x_c$ &  $\sqrt{1-{x_c}^2}$ & 0 & $\frac{3}{3-\gamma}$ & 
$|x_c|\leq\frac{1}{\sqrt{2}}$, { $x_c\neq 0$} 
   & 
$0$ &  $-\sqrt{6}\mu x_c$
         &  $-3\left(1-\sqrt{1-{x_c}^2}\right)$  &  
$-3\left(1+\sqrt{1-{x_c}^2}\right)$   & stable      &  
$1-2{x_c}^2$   &   $-1$        \\
   &&&&& with $\gamma<0$ &&&&& for $\mu x_c >0$ &&  \\
            &&&&&&&&&  &   &&   \\     
		 \hline\hline
	\end{tabular}%
 }
	\caption{The critical points for the case of  
 $V (\phi) = \frac{1}{8 \beta} \left(1 - e^{- \kappa \mu\; 
\phi}\right)^2$ and interaction rate $Q=\Gamma 
\rho_m$,
alongside their eigenvalues and their existence and stability conditions. 
Additionally, we present 
the corresponding values of the  dark-energy density parameter
$\Omega_\phi$ and the total equation-of-state parameter $w_{tot}$.  }
	\label{first-table-potIII-modelII}
\end{table*}

\begin{itemize}
    \item  Point $F_1$ exists always, and it corresponds to a dark energy 
dominated accelerating solution. Moreover, it is a stable node for 
$-\alpha<3+4\mu^2$ and 
hence  it can describe the late-time universe.  
    
\item The critical point $F_2$ corresponds to accelerating solutions 
for $\alpha<-1$, where $w_{tot}<-\frac{1}{3}$. We mention  that  at this 
  point  we have $\Omega_m/\Omega_\phi\neq 0$ corresponding to 
accelerating scaling solution, which is also stable in its stability region 
as shown in Table \ref{first-table-potIII-modelI} and in         Fig. 
\ref{fig1-model-III}, and thus it can 
alleviate the coincidence problem. Finally, one 
can see that for $\mu^2=-(\alpha+3)^2/4\alpha$ point $F_1$  corresponds to
completely matter dominated solutions.     

   \item  The curve of critical points $F_3$   has 
  $\lambda=0$, and it was absent for the exponential and hyperbolic 
potentials. In general these points  correspond to scaling 
  accelerating solutions, while for $x_c=\pm\frac{1}{\sqrt{2}}$  they 
correspond to  completely matter-dominated  points. Since, one eigenvalue is 
zero and the other two are negative,  they are normally 
hyperbolic critical points. By applying the   center manifold theorem  
\cite{Leon:2009dt} we deduce that these critical points behave as  
stable points for $\mu x_c >0$. Thus, these points can alleviate the 
coincidence problem for the model parameter $\alpha=-3$. 
   
\end{itemize}

Finally, from the critical point analysis at infinity (see 
Appendix \ref{appendix-5}), we deduce that the phase space domain $\mathbb{D}$ 
includes four such critical points, namely
$E_1^{\pm\infty}\left(\pm\frac{1}{\sqrt{2}},\frac{1}{\sqrt{2}},0,0\right)$, and 
$E_2^{\pm\infty}\left(\pm\sqrt{\frac{2}{7}},\sqrt{\frac{2}{7}},\mp\sqrt{\frac{3}
{7}},0\right)$. The first two points are unstable, while the last two are 
saddles. At these critical points, it follows that $q\longrightarrow-\infty$.

\subsubsection{Model II: $Q=\Gamma \rho_m$}

In this case, using   the dimensionless variables $x$, $y$, $\lambda$, $z$  
 of (\ref{dim-variables}), (\ref{dim-variable-z}) and
(\ref{dim-variable-lamda}), we result to the autonomous system
\begin{eqnarray}
x' &=& -3x-\frac{\sqrt{6}}{2}\lambda y^2+\frac{3}{2}x \left(1-x^2-y^2\right) 
\nonumber \\ && -\frac{\gamma z 
\left(1+x^2-y^2\right)}{2x(1-z)},\label{aut-sys-potIII-modelII-1}\end{eqnarray}
\begin{eqnarray}
y' &=& -\frac{\sqrt{6}}{2}\lambda x y+\frac{3}{2}y\left(1-x^2- y^2\right), 
\label{aut-sys-potIII-modelII-2}\end{eqnarray}
\begin{eqnarray}
 \lambda' &=&  \frac{\sqrt{6}}{2}\left(\lambda^2-2\mu\lambda \right)x, 
\label{aut-sys-potIII-modelII-3}
\end{eqnarray}
\begin{eqnarray}
z' &=& 
\frac{3}{2}z(1-z)\left(1-x^2-y^2\right),\label{aut-sys-potIII-modelII-4}
\end{eqnarray}
where $\gamma=\Gamma/H_0$. The 
above autonomous system is invariant under $y\longrightarrow-y$, and the phase 
space domain  is defined as  
$\mathbb{D}=\left\{ (x,y,\lambda,z)\in \mathbb{R}^4:  0\leq-x^2+y^2\leq1, 
y\geq 0,0\leq z\leq1\right\}$.
The dynamical system has two singular surfaces, namely $x=0$ and 
$z=1$, in which it is ill-defined. The properties and features of the critical 
points are summarized in  Table \ref{first-table-potIII-modelII}. They are the 
following.

\begin{table*}[]
\centering
\resizebox{1.0\textwidth}{!}{%
	\begin{tabular}{c c c c c c c c c}\hline\hline
          
		Point & $x$ & $v$ & $w$ & $\xi$ & Existence & Stability & 
$\Omega_\phi$ & $w_{tot}$\\ \hline
   &&&&&&&&  \\
        $I_1$ & $\frac{2}{3}$ & 0 & 0 & $\frac{\sqrt{13}}{3}$ & 
$\forall\alpha$ & stable for $\alpha>-\frac{17}{3}$ & 1 & $-\frac{17}{9}$\\
		
   &&&&&&&&  \\
         $I_2$ & $-\frac{\alpha+3}{4}$ & 0 & 0 & 
$\frac{\sqrt{-27-34\alpha-3\alpha^2}}{4\sqrt{3}}$ & 
$\forall$$\alpha\in\left[\frac{-13-\sqrt{88}}{3},-\frac{17}{3}\right]$ & stable 
if 
$\alpha\in\left[\frac{-13-\sqrt{88}}{3},-\frac{17}{3}\right]$ & 
$-\frac{3\alpha^2+26\alpha+27}{24}$ & $\frac{\alpha}{3}$\\
     & & & &  & $\cup\left(-3,\frac{-13+\sqrt{88}}{3}\right]$ & otherwise 
saddle 
or 
unstable & &  \\
      &&&&&&&&  \\
      $I_3$ & $\frac{1}{3}$ & 0 & $\frac{\sqrt{10}}{3}$ & 0 & 
$\forall\alpha$ 
& always saddle & 1 & $-\frac{11}{9}$\\
	 &&&&&&&&  \\
      $I_4$ & $-\frac{\alpha+3}{2}$ & 0 & 
$\frac{\sqrt{-27-22\alpha-3\alpha^2}}{2\sqrt{3}}$ & 0 & 
$\forall$$\alpha\in\left[\frac{-10-\sqrt{19}}{3},-\frac{11}{3}\right]$ & always 
saddle & $-\frac{3\alpha^2+20\alpha+27}{6}$ & $\frac{\alpha}{3}$\\
      &&&&& $\cup \left(-3,\frac{-10+\sqrt{19}}{3}\right]$ &&& \\
     &&&&&&&&  \\
        \hline\hline
	\end{tabular}%
 }
	\caption{The critical points for the case of  
  $V(\phi)=c_0+c_1 e^{\sqrt{\frac{2\kappa^2}{3}}\phi}+c_2 
e^{2\sqrt{\frac{2\kappa^2}{3}}\phi}$ and interaction rate $Q=\alpha H \rho_m$,
alongside their existence and stability conditions. Furthermore, we present 
the corresponding values of the  dark-energy density parameter
$\Omega_\phi$ and the total equation-of-state parameter $w_{tot}$.    }
	\label{first-table-potIV-modelI}
\end{table*}

\begin{itemize}

\item  The critical point $G_1$ exists always, corresponding to
dark-energy dominated super-accelerating solutions,  
with $H\longrightarrow\infty$, and it is always stable node. Thus, $G_1$ is a 
candidate for describing the late-time accelerated universe.

\item  Point $G_2$ exists always, corresponding to
dark-energy dominated  accelerating solutions,  with $H\longrightarrow0$, and 
it is a saddle point.

  \item The curves of critical points $G_3$ exist  for 
$|x_c|\leq\frac{1}{\sqrt{2}}$ with $\gamma<0$.  They are normally hyperbolic, 
and by applying the center manifold theorem in order to 
  determine their stability \cite{Leon:2009dt} we deduce that      $G_3$ are 
stable for $\mu x_c>0$. 
    Note that in general, for  $|x_c| < 
\frac{1}{\sqrt{2}}$,  these points have 
 $0 < \Omega_\phi<1$, and thus  they   correspond to a late-time stable 
accelerating solution with $\Omega_m/\Omega_{\phi} \neq 0$, namely with a 
scaling nature  that can alleviate the coincidence problem. 
Finally, in the limiting case  where  $x_c=\pm\frac{1}{\sqrt{2}}$ they 
lead to $\Omega_\phi=0$ and thus to $\Omega_m=1$, corresponding to  
  matter-dominated solutions.

\end{itemize}

Lastly, the system encounters two critical points at infinity, namely  
$F^{+\infty}\Bigl\{\left(y_1,y_2,y_3,y_4,y_5\right)\in\mathbb{S}^4:y_5=0,y_4=0,
y_1^2=y_2^2, y_1>0,y_2>0\Bigr\}$ and 
$F^{-\infty}\Bigl\{\left(y_1,y_2,y_3,y_4,y_5\right)\in\mathbb{S}^4:y_5=0,y_4=0,
y_1^2=y_2^2, y_1<0,y_2>0\Bigr\}$, where $q\longrightarrow-\infty$. These critical points exhibit
 unstable behavior. The details of the analysis are presented in Appendix 
\ref{appendix-6}.
 
  In summary,   under the local interaction rate  the potential $V 
(\phi) = \frac{1}{8 \beta} \left(1 - e^{- \kappa \mu\; 
\phi}\right)^2$   is able to accept    
late-time stable scaling accelerating solutions, which are not possible for 
the previous potentials, and that is why they were not studied in the 
literature. This is a novel result of the present work.

\subsection{Potential: $V(\phi)=c_0+c_1 e^{\sqrt{\frac{2\kappa^2}{3}}\phi}+c_2 
e^{2\sqrt{\frac{2\kappa^2}{3}}\phi}$}

In this subsection  we focus on the last  potential function 
(\ref{gen-exp-potential}).
In order to obtain the autonomous form for this potential case  we introduce 
three new dimensionless variables $v$, $w$, $\xi$ as 
\begin{equation} 
\label{new-dim-var}
    v=\frac{\kappa\sqrt{c_0}}{\sqrt{3}H}, ~w=\frac{\kappa\sqrt{c_1 
e^{\sqrt{\frac{2\kappa^2}{3}}\phi}}}{\sqrt{3}H},~\xi=\frac{\kappa\sqrt{c_2 
e^{2\sqrt{\frac{2\kappa^2}{3}}\phi}}}{\sqrt{3}H}.
\end{equation}
In terms of these variables, as well as    $x$ 
defined in (\ref{dim-variables}), the various cosmological parameters become 
\begin{eqnarray}
&& \Omega_m = 1+x^2-v^2-w^2-\xi^2, \nonumber \\ 
&& \Omega_\phi = -x^2+v^2+w^2+\xi^2, \nonumber \\
&& w_{tot} = -x^2-v^2-w^2-\xi^2, \nonumber \\ 
&& q = \frac{1}{2}\left[1-3(x^2+v^2+w^2+\xi^2)\right].  
\end{eqnarray}

\subsubsection{Model I: $Q=\alpha H \rho_m$}

Using the above dimensionless variables   we   obtain the autonomous 
system as
\begin{eqnarray}
    x'&=&-\frac{3}{2}x \left(1+x^2+v^2+w^2+\xi^2\right)+w^2+2\xi^2 \nonumber \\ 
&& 
-\frac{\alpha\left(1+x^2-v^2-w^2-\xi^2\right)}{2x},\label{aut-sys-potIV-modelI-1}  \\
v'&=&\frac{3}{2}v\left(1-x^2-v^2-w^2-\xi^2\right),\label{aut-sys-potIV-modelI-2}
\\ 
  w'&=&w\left[x+\frac{3}{2}\left(1-x^2-v^2-w^2-\xi^2\right)\right],
  \label{aut-sys-potIV-modelI-3}
\end{eqnarray}
\begin{eqnarray}  
\xi'&=&\xi\left[2x+\frac{3}{2}\left(1-x^2-v^2-w^2-\xi^2\right)\right].
\label{aut-sys-potIV-modelI-4}
\end{eqnarray}
This system is  invariant under the transformations $v\longrightarrow-v$, 
$w\longrightarrow-w$ and $\xi\longrightarrow-\xi$ separately, and thus   the 
physical region   is defined as 
\begin{eqnarray}\label{phy-reg-potIV-modelI}
    \mathbb{D}&=&\bigg\{ (x,v,w,\xi)\in \mathbb{R}^4: 
0\leq-x^2+v^2+w^2+\xi^2\leq1, \nonumber\\ && v\geq0, w\geq0, \xi\geq0 \bigg\}.
\end{eqnarray}
The system exhibits a singular surface at $x=0$, where it becomes 
ill-defined.
 The critical points, the existence and
stability conditions as well as the corresponding values of  
$\Omega_\phi$ and $w_{tot}$ are 
summarized in Table \ref{first-table-potIV-modelI}, while the corresponding 
eigenvalues are displayed in Table \ref{second-table-potIV-modelI}.
For this scenario we have the following critical points.

\begin{table}[!]
\centering
 {%
	\begin{tabular}{c c c c c c}\hline\hline
          
		Point & $E_1$ & $E_2$ & $E_3$ & $E_4$  \\ \hline
    &&&&  \\
		$I_1$ & $-\frac{17}{3}-\alpha$ & $-\frac{4}{3}$ &  $-\frac{2}{3}$ &  
$-\frac{13}{3}$  \\
   &&&&  \\
        $I_2$ & $\frac{3+\alpha}{4}$ & $\frac{3+\alpha}{2}$ &  
$\frac{T'-\sqrt{V'}}{12(3+\alpha)}$ &  $\frac{T'+\sqrt{V'}}{12(3+\alpha)}$ \\
   &&&&  \\
         $I_3$ & $-\frac{11}{3}-\alpha$ & $-\frac{1}{3}$ &  $-\frac{10}{3}$ & 
$\frac{1}{3}$ \\
   &&&&  \\
        $I_4$ & $\frac{3+\alpha}{2}$ & $-\frac{3+\alpha}{2}$ &  
$\frac{R'-\sqrt{S'}}{12(3+\alpha)}$ &  $\frac{R'+\sqrt{S'}}{12(3+\alpha)}$ \\
   &&&&  \\ 
            \hline\hline
	\end{tabular}%
 }
	\caption{The critical points and their associated  eigenvalues of the 
linearised (Jacobian) matrix, for the case  of  scalar-field potential 
 $V(\phi)=c_0+c_1 
e^{\sqrt{\frac{2\kappa^2}{3}}\phi}+c_2 e^{2\sqrt{\frac{2\kappa^2}{3}}\phi}$ and 
interaction rate  $Q=\alpha H \rho_m$. We 
have defined      $T'=-27+34\alpha+9\alpha^2$, 
$V'=-11664-27891\alpha-16700\alpha^2-4362\alpha^3-540\alpha^4-27\alpha^5$, 
$R'=-27+22\alpha+9\alpha^2$ and 
$S'=-31347-57456\alpha-37514\alpha^2-11580\alpha^3-1755\alpha^4-108\alpha^5$.  }
	\label{second-table-potIV-modelI}
\end{table}

\begin{itemize}

\item The critical point $I_1$ exists for all   $\alpha$, and it 
corresponds to a dark-energy dominated super-accelerated
solution.   It  is stable in the region 
$\alpha>-\frac{17}{3}$ and thus it can describe the 
late-time phase of the universe.
    
\item Point  $I_2$ in its stability region has $0\leq\Omega_\phi\leq1$, and 
thus $0\leq\Omega_m\leq1$. Hence, since $\Omega_\phi$ and $\Omega_m$ can be of 
the same order, this point can alleviate the   coincidence problem 
\cite{Amendola:1999er,Chimento:2003iea,Cai:2004dk,Pavon:2005yx,Huey:2004qv, 
Hu:2006ar,delCampo:2008sr,delCampo:2008jx}.  Moreover, for  
$\alpha=\frac{-13-\sqrt{88}}{3}$ or 
$\alpha=\frac{-33+\sqrt{88}}{3}$, it has  $\Omega_m=1$, and in this case it 
corresponds to a matter-dominated solution.
 
  \item Point $I_3$ exists always, it is saddle, and it corresponds to a 
dark-energy dominated 
super-accelerating solution.

\begin{table*}[] 
\centering
 {%
	\begin{tabular}{c c c c c c c c c c}\hline\hline
          
		Point & $x$ & $v$ & $w$ & $\xi$ & $z$ & Existence & Stability & 
$\Omega_\phi$ & $w_{tot}$\\ \hline
   &&&&&&&&&  \\
        $J_1$ & $\frac{1}{3}$ & 0 & $\frac{\sqrt{10}}{3}$ & 0 & 0 & 
$\forall\gamma$ & always saddle & 1 & $-\frac{11}{9}$\\
   &&&&&&&&&  \\
        $J_2$ & $\frac{2}{3}$ & 0 & 0 & $\frac{\sqrt{13}}{3}$ & 0 & 
$\forall\gamma$ & always stable & 1 & $-\frac{17}{9}$\\
   &&&&&&&&&  \\
          $J_3$ & $\frac{1}{3}$ & 0 & $\frac{\sqrt{10}}{3}$ & 0 & 1 & 
$\forall\gamma$ & always saddle & 1 & $-\frac{11}{9}$\\
   &&&&&&&&&  \\
          $J_4$ & $\frac{2}{3}$ & 0 & 0 & $\frac{\sqrt{13}}{3}$ & 1 & 
$\forall\gamma$ & always saddle & 1 & $-\frac{17}{9}$\\
   &&&&&&&&&  \\
          $J_5$ & $x_c$ & $\sqrt{1-{x_c}^2}$ & 0 & 0 & 
$\ \ \ \frac{3}{3-\gamma}\ \ \ $ & $\ \ \ 0<|x_c|\leq\frac{1}{\sqrt{2}}\ \ \ $
  & stable for $-\frac{1}{\sqrt{2}}\leq x_c<0\ $ & $\ \ \ 1-2{x_c}^2\ \ \ $ & $ 
-1$\\
   &&&&&& with $\gamma<0$ &  &&     \\
               \hline\hline
	\end{tabular}%
 }
	\caption{The critical points for the case of  
  $V(\phi)=c_0+c_1 e^{\sqrt{\frac{2\kappa^2}{3}}\phi}+c_2 
e^{2\sqrt{\frac{2\kappa^2}{3}}\phi}$ and interaction rate $Q=\Gamma\rho_m$,
alongside their existence and stability conditions. Furthermore, we present 
the corresponding values of the  dark-energy density parameter
$\Omega_\phi$ and the total equation-of-state parameter $w_{tot}$.     }
	\label{first-table-potIV-modelII}
\end{table*}

  \item Point $I_4$  has $0\leq\Omega_\phi\leq1$  
and   $0\leq\Omega_m\leq1$, and corresponds to accelerating solutions. However, 
 since it is saddle  it cannot offer an  alleviation for the coincidence 
problem. Finally, in the limiting case where 
  $\alpha=\frac{-10\pm\sqrt{19}}{3}$  we have  $\Omega_{m} 
= 1$, corresponding to a matter-dominated solution.

\end{itemize}

Finally, we observe that the system possesses
two critical points at infinity, namely
$G^{\pm\infty}\Bigl\{\left(y_1,y_2,y_3,y_4,y_5\right)\in\mathbb{S}^4:y_5=0,y_4>0
, y_3>0, y_2>0, y_1=\pm\frac{1}{\sqrt{2}}\Bigr\}$, exhibiting unstable 
behavior and having 
$q\longrightarrow-\infty$. The 
details of the analysis are presented in Appendix \ref{appendix-7}.

\subsubsection{Model II: $Q=\Gamma\rho_m$}

 In this case, using the dimensionless variables $v$, $w$, $\xi$ defined in 
(\ref{new-dim-var}), the variable $z$ defined in (\ref{dim-variable-z}) and 
the dimensionless variable $x$ from (\ref{dim-variables}), the autonomous 
system can be written as  
\begin{eqnarray}
    x'&=&-\frac{3}{2}x\left(1+x^2+v^2+w^2+\xi^2\right)+w^2+2\xi^2  \nonumber \\ 
&& -\frac{\gamma\left(1+x^2-v^2-w^2-\xi^2\right)z}{2x(1-z)}, 
\label{aut-sys-potIV-modelII-1} \end{eqnarray}
\begin{eqnarray}  
v'&=&\frac{3}{2}v\left(1-x^2-v^2-w^2-\xi^2\right),\label{aut-sys-potIV-modelII-2
}\\ 
    w'&=&w\left[x+\frac{3}{2}\left(1-x^2-v^2-w^2-\xi^2\right)\right],\label{
aut-sys-potIV-modelII-3}\end{eqnarray}
\begin{eqnarray} 
\xi'&=&\xi\left[2x+\frac{3}{2}\left(1-x^2-v^2-w^2-\xi^2\right)\right],\label{ 
aut-sys-potIV-modelII-4}\end{eqnarray}
\begin{eqnarray} 
z'&=&\frac{3}{2}z(1-z)\left(1-x^2-v^2-w^2-\xi^2\right),\label{aut-sys-potIV-modelII-5} \end{eqnarray}
where   $\gamma=\Gamma/H_0$ is the dimensionless coupling parameter.
The system is  invariant under the transformations $v\longrightarrow-v$, 
$w\longrightarrow-w$ and $\xi\longrightarrow-\xi$ separately, and thus   the 
physical region   is defined as 
 \begin{eqnarray}\label{phy-reg-potIV-modelII}
   \mathbb{D}&=&\bigg\{ (x,v,w,\xi,z)\in \mathbb{R}^5: 0\leq-x^2+v^2+w^2 
+\xi^2\leq1, \nonumber \\  && v\geq0, w\geq0, \xi\geq0, 0\leq z \leq1 \bigg\}.
\end{eqnarray}
There are two singular surfaces in the dynamical system, at $x=0$ 
and $z=1$ respectively, at which the system is ill-defined.

 The critical points, the existence and
stability conditions as well as the corresponding values of  
$\Omega_\phi$ and $w_{tot}$ are 
summarized in Table \ref{first-table-potIV-modelII}, while the corresponding 
eigenvalues are displayed in    Table \ref{second-table-potIV-modelII}.
For this scenario we have the following critical points.

\begin{table}[ht]
\resizebox{0.48\textwidth}{!}{%
	\begin{tabular}{c c c c c c c }\hline\hline          
		Point & $E_1$ & $E_2$ & $E_3$ & $E_4$ & $E_5$ \\ \hline
   &&&&&  \\
		$J_1$ & $-\frac{11}{3}$ & $-\frac{1}{3}$ & $-\frac{10}{3}$ & 
$\frac{1}{3}$ & $-\frac{1}{3}$ \\
   &&&&&  \\
        $J_2$ & $-\frac{17}{3}$ & $-\frac{4}{3}$ & $-\frac{2}{3}$ & 
$-\frac{13}{3}$ & $-\frac{4}{3}$ \\
   &&&&&  \\
         $J_3$ & $-sgn(\gamma)\infty$ & $-\frac{1}{3}$ & $-\frac{10}{3}$ & 
$\frac{1}{3}$ & $\frac{1}{3}$ \\
   &&&&&  \\
          $J_4$ & $-sgn(\gamma)\infty$ & $-\frac{4}{3}$ & $-\frac{2}{3}$ & 
$-\frac{13}{3}$ & $\frac{4}{3}$ \\
   &&&&&  \\
           $J_5$ & $0$ & $x_c$ & $2x_c$ & $-3(1+\sqrt{1-{x_c}^2})$ & 
$-3(1-\sqrt{1-{x_c}^2})$ \\
   &&&&&  \\      
            \hline\hline
	\end{tabular}
	}
	\caption{The critical points and their associated  eigenvalues of the 
linearised (Jacobian) matrix, for the case  of  scalar-field potential 
 $V(\phi)=c_0+c_1 
e^{\sqrt{\frac{2\kappa^2}{3}}\phi}+c_2 e^{2\sqrt{\frac{2\kappa^2}{3}}\phi}$ and 
interaction rate  $Q=\Gamma \rho_m$.    }
	\label{second-table-potIV-modelII}
\end{table}

\begin{itemize}

\item The critical point $J_1$ does not depend on the parameter 
$\gamma$, it exists always and it is saddle, corresponding to 
super-accelerated dark-energy dominated solution, with 
$H\longrightarrow\infty$. 
  
\item Point $J_2$ exists always and it is 
stable, corresponding to super-accelerated dark-energy dominated 
solution, with  $H\longrightarrow\infty$. 
Thus, $J_2$ can describe the late-time evolution of the universe.

\item Point $J_3$ exists always, it is saddle, and it also corresponds 
to super-accelerated dark-energy dominated solution, but  with  
$H\longrightarrow0$. 

\begin{figure}[!]
	\includegraphics[width=0.47\textwidth]{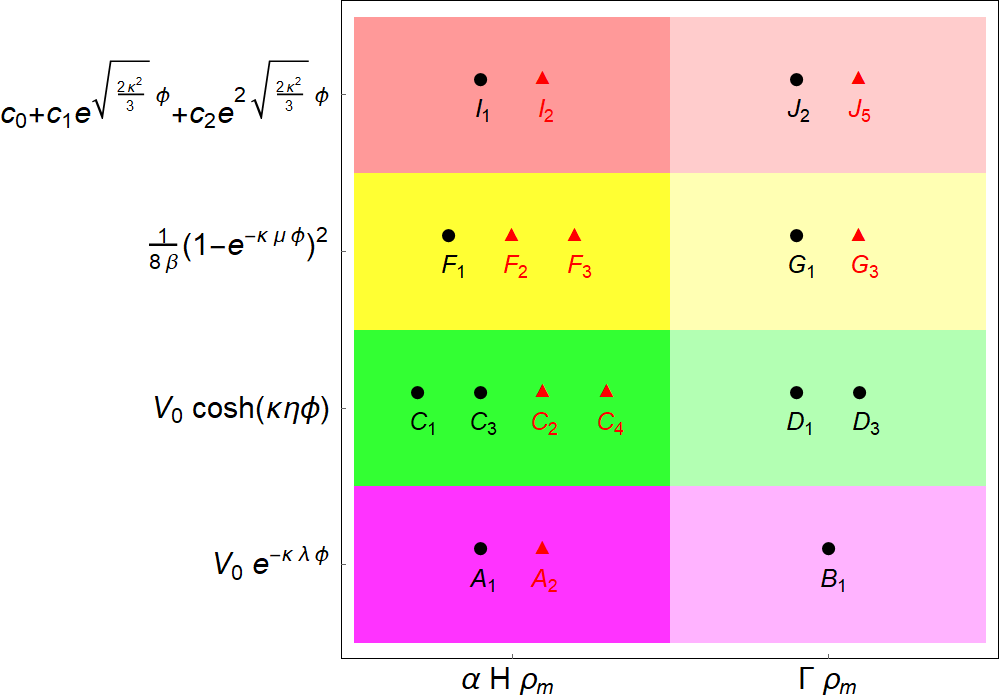}
	\caption{{\it{Summary plot of the phase-space behavior and the critical 
points for the interaction models $Q=\alpha H \rho_m$ and $Q=\Gamma \rho_m$, 
considering all the examined potentials. Each row corresponds to a specific 
potential and color pattern (magenta, green, yellow, pink), and each column to a
specific interaction model and  shade pattern
 (dark shade, light shade). We only focus on late-time attractors. Black 
circles correspond to the late-time dark-energy dominated accelerating 
attractors, while red triangles indicate late-time accelerating scaling 
attractors. 
For the global interaction model $Q=\alpha H \rho_m$, all   potentials can 
provide accelerating scaling attractors, while for local interaction rate 
$Q=\Gamma\rho_m$, only two potentials, namely  $V (\phi) = \frac{1}{8 \beta} 
\left(1 - e^{- \kappa \mu\; 
\phi}\right)^2$ and $V(\phi)=c_0+c_1 e^{\sqrt{\frac{2\kappa^2}{3}}\phi}+c_2 
e^{2\sqrt{\frac{2\kappa^2}{3}}\phi}$ can produce accelerating scaling 
attractors. Finally,  note that all   scaling 
solutions can 
transform to   matter-dominated solution in the limiting case of the parameter 
region (see text).
 }}} 
	\label{summary-plot}
\end{figure}

 \item Point $J_4$ exist always,  it is saddle, and   it also   
corresponds to a super-accelerated dark-energy dominated solution, with  
$H\longrightarrow0$.  

 \item The curve of critical points $J_5$ corresponds to accelerating 
solutions. Since    these points are  normally hyperbolic, by applying the 
center manifold 
theorem we deduce that  $J_5$ points are stable for $-\frac{1}{\sqrt{2}}\leq 
x_c<0$. Additionally, note that for this   curve of critical points we 
have   $0\leq\Omega_\phi\leq1$ 
and $0\leq\Omega_m\leq1$, and consequently  
$\Omega_m/\Omega_{\phi} \neq 0$, and hence $J_5$ can alleviate the 
coincidence problem. Finally, in the 
case  $x_c=\pm\frac{1}{\sqrt{2}}$ one has  $\Omega_{m} = 1$, corresponding to a 
completely 
matter-dominated solution.  
    
\end{itemize}

Lastly, the Poincar\'{e} compactification of the system results in two critical points at infinity, given by, 
$J^{\pm\infty}\Bigl\{\left(y_1,y_2,y_3,y_4,y_5,y_6\right)\in\mathbb{S}^5:y_6=0,
y_5=0,y_4>0, y_3>0, y_2>0, y_1=\pm\frac{1}{\sqrt{2}}\Bigr\}$, which exhibit 
unstable behavior, and having 
$q\longrightarrow-\infty$. The 
details are presented in Appendix \ref{appendix-8}.

\section{Summary and Conclusions}
\label{sec-conclusion}

Cosmological scenarios which include interactions between dark matter and 
dark energy sectors  have received  significant attention in the 
scientific community since they exhibit  many appealing features, such as the 
solution of the cosmic coincidence problem and the alleviation of $H_0$ and 
$S_8$ tensions. On the other hand, phantom cosmology has also attracted a 
significant amount of research, since it can efficiently describe the phantom 
crossing.

In the present work we performed a detailed investigation of interacting 
phantom cosmology, by applying the powerful method of dynamical system 
analysis, since this approach allows to extract information about the global 
features of the scenario, independently of the initial conditions or the 
specific evolution of the Universe. The novel ingredient of our work is the 
examination of new potentials, which   indeed lead to new results even for the 
same interaction forms previously studied in the literature. 

We considered two well-studied interaction forms, namely the global one   $Q = 
\alpha H 
\rho_m$ and the local one $Q = \Gamma \rho_m$. Additionally, we considered 
four widely-used potentials for the 
phantom scalar field, namely A) 
$V(\phi)=V_0 e^{- \kappa \lambda \phi}$, B) $V(\phi)=V_0
\cosh( \kappa \eta \phi)$, C) $V  (\phi) = \left(1 - e^{- \kappa \mu \; \phi}\right)^2/ 8 
\beta$ and
D) $V(\phi)=c_0+c_1 e^{\sqrt{2\kappa^2/3}\phi}+c_2 
e^{2\sqrt{2\kappa^2/3}\phi}$, where the last two are the new ones that have not 
been studied before in the framework of interacting phantom cosmology.
In all cases, we extracted the critical points and we 
examined their stability.

Our analysis extracted   saddle  matter-dominated points, stable  dark-energy 
dominated points, and scaling accelerating solutions,  that can attract the 
Universe at late times. 
As we discussed, some of the accelerating scaling attractors 
obtained for the last two potentials, when the local interaction is considered, 
are totally new. Furthermore, performing a detailed analysis at 
infinity it has been shown that there are no stable critical points at infinity, and thus they cannot 
attract the Universe at late times.
For completeness, in  Fig. \ref{summary-plot} we present the 
summary diagram that provides  the qualitative behavior of the examined 
interacting phantom scenarios for 
different potentials.

In summary, our analysis revealed that interacting phantom 
cosmology can describe the Universe evolution very efficiently, namely the 
transition from a matter-dominated  saddle  point
to a stable dark-energy dominated 
accelerated universe,  or a stable accelerating phase in which dark-energy and 
dark matter co-exist,  and hence alleviating the cosmic coincidence problem.  
These features make interacting phantom 
cosmology a good candidate for the description of nature.

\section{Acknowledgments}

The authors would like to thank an anonymous referee for   
important comments that significantly improved the quality of the manuscript.  
SH acknowledges  the financial support from the University Grants Commission 
(UGC), Govt. of India (NTA Ref. No: 201610019097). SP and TS acknowledge the 
financial support from the Department of Science and Technology (DST), Govt. of 
India under the Scheme   ``Fund for Improvement of S\&T Infrastructure (FIST)'' 
(File No. SR/FST/MS-I/2019/41). ENS acknowledges the contribution of the LISA 
CosWG, and of   COST 
Actions  CA18108  "Quantum Gravity Phenomenology in the multi-messenger 
approach''  and  CA21136" Addressing observational tensions in cosmology with 
systematics and fundamental physics (CosmoVerse)''.  

\begin{appendix}

\section{Potential $V(\phi)=V_0 e^{-\kappa\lambda\phi}$}

\subsection{Model I: $Q=\alpha H \rho_m$}\label{appendix-1}

The phase space domain of the autonomous system 
(\ref{aut-sys-potI-modelI-1})-(\ref{aut-sys-potI-modelI-2}) is not compact in 
the $x-y$ direction. Thus, one may expect that there are critical points at 
infinity. Since the autonomous system is ill-defined at $x=0$ we divide the 
phase-space domain $\mathbb{D}$ into two parts: the first one for $x>0$ and the 
second one for $x<0$, i.e. we remove $x=0$ singular line from our phase space 
and we resort to Poincar\'{e} compactification technique 
\cite{perko2013differential,dumortier2006qualitative,meiss2007differential, 
Bahamonde:2017ize} to determine the qualitative behavior of the critical points 
at infinity.

In the part of the phase space where $x>0$, we use a re-parametrization, i.e., we introduce a new time variable $\tau$,
defined by $dN=x d\tau$, and the system 
(\ref{aut-sys-potI-modelI-1})-(\ref{aut-sys-potI-modelI-2}) simplifies as
\begin{eqnarray}
\frac{dx}{d\tau} &=& -3x^2-\frac{\sqrt{6}}{2}\lambda x y^2+\frac{3}{2}x^2\left(1-x^2-y^2\right)  
\nonumber\\ && - 
\frac{\alpha}{2}\left(1+x^2-y^2\right),\label{reg-aut-sys-potI-modelI-1}\\
\frac{dy}{d\tau} &=& -\frac{\sqrt{6}}{2}\lambda x^2 y+\frac{3}{2}xy\left(1-x^2-y^2\right).  
\label{reg-aut-sys-potI-modelI-2}
\end{eqnarray}
The vector field associated with the above system  is defined as $(P,Q)$ where 
$P=-3x^2-\frac{\sqrt{6}}{2}\lambda x 
y^2+\frac{3}{2}x^2\left(1-x^2-y^2\right)-\frac{\alpha}{2}\left(1+x^2-y^2\right)$ 
and $Q=-\frac{\sqrt{6}}{2}\lambda x^2 y+\frac{3}{2}xy\left(1-x^2-y^2\right)$. 
Now, we project this vector field on the upper hemisphere $\mathbb{S}^{2+}$ 
using the central projection $y_1=\frac{x}{\Delta x}$, $y_2=\frac{y}{\Delta x}$ 
and $y_3=\frac{1}{\Delta x}$, where $\Delta x=\sqrt{1+x^2+y^2}$. The stability 
analysis is performed in three different local charts, namely 
$\left(U_i,\phi_i\right)$, where $i=1,2,3$, 
$U_i=\left\{(y_1,y_2,y_3)\in\mathbb{S}^2:y_i>0\right\}$, and the local map  
$\phi_i:U_i\longrightarrow\mathbb{R}^2$ is defined as 
$\phi_i\left(y_1,y_2,y_3\right)=\left(\frac{y_n}{y_i},\frac{y_m}{y_i}\right)$ 
for $n<m$ and $n,m\neq i$.

In the first chart, we set $\phi_1\left(y_1,y_2,y_3\right)=(u,v)$, reducing the 
system (\ref{reg-aut-sys-potI-modelI-1})-(\ref{reg-aut-sys-potI-modelI-2}) as 
\begin{eqnarray}
\frac{du}{d\tau} &=& 3uv^2+\frac{\sqrt{6}}{2}\lambda u v \left(u^2-1\right) \nonumber\\ &&
-\frac{\alpha}{2} u v^2\left(1-u^2+v^2\right)  
,\label{chart1-modelI-1}\\
\frac{dv}{d\tau} &=& 3v^3+\frac{\sqrt{6}}{2}\lambda u^2 v^2+\frac{3}{2}v\left(1+u^2-v^2\right) \nonumber\\ 
&& \frac{\alpha}{2}v^3\left(1-u^2+v^2\right).  
\label{chart1-modelI-2}
\end{eqnarray}
Since, $\frac{du}{d\tau}=0$, $\frac{dv}{d\tau}=0$ at $v=0$, the critical points 
at infinity are $\{(u,0):u\in\mathbb{R}\}$, which correspond to the equator 
circle of $\mathbb{S}^2$ with $y_1>0$. For our phase-space domain, the point at 
infinity is the point 
$A^{+\infty}\left(\frac{1}{\sqrt{2}},\frac{1}{\sqrt{2}},0\right)$, which 
further indicates the point $(1,0)$ in the $u-v$ plane. The eigenvalues at   
point $(1,0)$ are $3$ and $0$, concluding that it is an unstable point i.e. 
$A^{+\infty}$ is unstable.

For the second chart  we consider $\phi_2\left(y_1,y_2,y_3\right)=(u,v)$, which 
leads to the system
\begin{eqnarray}
\frac{du}{d\tau} &=& -3u^2v^2+\frac{\sqrt{6}}{2}\lambda u v \left(u^2-1\right) \nonumber\\ &&
-\frac{\alpha}{2} v^2\left(u^2+v^2-1\right)  
,\label{chart2-modelI-1}\\
\frac{dv}{d\tau} &=&\frac{\sqrt{6}}{2}\lambda u^2 v^2-\frac{3}{2}u v\left(v^2-u^2-1\right).  
\label{chart2-modelI-2}
\end{eqnarray}
Note that here the curve 
$\{(u,0):u\in\mathbb{R}\}$ also corresponds to points at infinity, which is the 
equator circle of $\mathbb{S}^2$ with $y_2>0$. Once again, the point 
$A^{+\infty}\left(\frac{1}{\sqrt{2}},\frac{1}{\sqrt{2}},0\right)$ on the equator 
circle is the point at infinity for our physical domain. The point $A^{+\infty}$ 
associates the point $(1,0)$ on the $u-v$ plane. The point $(1,0)$ is also 
unstable since one eigenvalue is $2$ and the other is zero. Thus, $A^{+\infty}$ 
exhibits unstable behavior.

In the third chart, if we  assume $\phi_2\left(y_1,y_2,y_3\right)=(u,v)$, the 
dynamical system 
(\ref{reg-aut-sys-potI-modelI-1})-(\ref{reg-aut-sys-potI-modelI-2})   becomes
\begin{eqnarray}
\frac{du}{d\tau} &=& P(u,v)  
,\label{u-chart3-modelI-1}\\
\frac{dv}{d\tau} &=&Q(u,v),  
\label{chart3-modelI-2}
\end{eqnarray}
which is identical to the system 
(\ref{reg-aut-sys-potI-modelI-1})-(\ref{reg-aut-sys-potI-modelI-2}) and 
hence it cannot provide any new solutions at infinity.

When we deal with the phase-space domain $\mathbb{D}$ with $x<0$, we use the re-parametrization $dN=-xd\tau$, transforming the system 
(\ref{aut-sys-potI-modelI-1})-(\ref{aut-sys-potI-modelI-2}) as
\begin{eqnarray}
\frac{dx}{d\tau} &=& 3x^2+\frac{\sqrt{6}}{2}\lambda x y^2-\frac{3}{2}x^2\left(1-x^2-y^2\right)  
\nonumber\\ && + 
\frac{\alpha}{2}\left(1+x^2-y^2\right),\label{n-reg-aut-sys-potI-modelI-1}\\
\frac{dy}{d\tau} &=& \frac{\sqrt{6}}{2}\lambda x^2 y-\frac{3}{2}xy\left(1-x^2-y^2\right).  
\label{n-reg-aut-sys-potI-modelI-2}
\end{eqnarray}
By applying the same procedure used in the previous paragraphs, in this case  we 
also extract a critical point at infinity, namely
$A^{-\infty}\left(-\frac{1}{\sqrt{2}},\frac{1}{\sqrt{2}},0\right)$, which 
exhibits unstable behavior.

In summary, the original system 
(\ref{aut-sys-potI-modelI-1})-(\ref{aut-sys-potI-modelI-2}) has two points at 
infinity, namely  
$A^{\pm\infty}\left(\pm\frac{1}{\sqrt{2}},\frac{1}{\sqrt{2}},0\right)$, which 
correspond to unstable solutions, and thus they cannot attract the Universe at 
late times.

\subsection{Model II: $Q=\Gamma \rho_m$}\label{appendix-2}

The dynamical system 
(\ref{aut-sys-potI-modelII-1})-(\ref{aut-sys-potI-modelII-3}) exhibits a 
singularity at $x=0$, making the system ill-defined there. Thus, in the part 
of the domain $\mathbb{D}$ where $x>0$, we utilize the re-parametrization $dN=x(1-z)d\tau$, simplifying the system 
(\ref{aut-sys-potI-modelII-1})-(\ref{aut-sys-potI-modelII-3}) to the form   
\begin{eqnarray}
\frac{dx}{d\tau}&=&-3x^2(1-z)+\frac{3}{2}x^2\left(1-x^2-y^2\right)(1-z)  
\nonumber \\ && -\frac{\sqrt{6}}{2}\lambda x y^2(1-z)-\frac{\gamma z 
\left(1+x^2-y^2\right)}{2},\label{reg-aut-sys-potI-modelII-1}\\
\frac{dy}{d\tau}&=& \left[-\frac{\sqrt{6}}{2}\lambda x^2 y+\frac{3}{2} x y\left(1-x^2- 
y^2\right)\right](1-z),\ \ \\ 
\label{reg-aut-sys-potII-modelII-2}
\frac{dz}{d\tau}&=& \frac{3}{2} x z(1-z)^2\left(1-x^2-y^2\right).\label{reg-aut-sys-potI-modelII-3}
\end{eqnarray}
The vector field related to the system mentioned above is defined as  
$(P,Q,R)$, in which $P=-3x^2(1-z) -\frac{\sqrt{6}}{2}\lambda x 
y^2(1-z)+\frac{3}{2}x^2\left(1-x^2-y^2\right)(1-z)-\frac{\gamma z 
\left(1+x^2-y^2\right)}{2}$, $Q= \left[-\frac{\sqrt{6}}{2}\lambda x^2 
y+\frac{3}{2} x y\left(1-x^2- y^2\right)\right](1-z)$ and $R=\frac{3}{2} x 
z(1-z)^2\left(1-x^2-y^2\right)$. Now, we proceed by projecting this vector  
field onto the upper three-sphere $\mathbb{S}^{3+}$ through central projection 
$y_1=\frac{x}{\Delta x}$, $y_2=\frac{y}{\Delta x}$, $y_3=\frac{z}{\Delta x}$ and 
$y_4=\frac{1}{\Delta x}$, where $\Delta x=\sqrt{1+x^2+y^2+z^2}$. The stability 
analysis is conducted in four distinct local charts, denoted as 
$\left(U_i,\phi_i\right)$, where $i=1,2,3,4$ and 
$U_i=\left\{\left(y_1,y_2,y_3,y_4\right)\in\mathbb{S}^3:y_i>0\right\}$, while 
the local map $\phi_i:U_i\longrightarrow\mathbb{R}^3$ is defined by 
$\phi_i\left(y_1,y_2,y_3,y_4\right)=\left(\frac{y_n}{y_i},\frac{y_m}{y_i},\frac{
y_p}{y_i}\right)$ for $n<m<p$ and $n,m,p\neq i$.

For the first chart, taking $\phi_1\left(y_1,y_2,y_3,y_3\right)=(u,v,w)$,  the 
dynamical system 
(\ref{reg-aut-sys-potI-modelII-1})-(\ref{reg-aut-sys-potI-modelII-3}) becomes
\begin{eqnarray}
\frac{du}{d\tau}&=&3uw^3(w-v)+\frac{\gamma u v w^3(w^2+1-u^2)}{2}  
\nonumber \\ && +\frac{\sqrt{6}}{2} u w^2(w-v)(u^2-1),\label{first-chart-potI-modelII-1}\\
\frac{dv}{d\tau} &=&\frac{3}{2}vw(w-v)\left(1+u^2+w^2\right) \nonumber \\&& 
+\frac{\sqrt{6}}{2} \lambda u^2vw^2(w-v)+\frac{\gamma v^2 
w^3\left(w^2+1-u^2\right)}{2} \nonumber \\
&&+\frac{3}{2}v(w-v)^2\left(w^2-1-u^2\right),\\
\label{first-chart-potI-modelII-2}
\frac{dw}{d\tau} &=& \frac{\sqrt{6}}{2}\lambda u^2 w^3 (w-v)+\frac{\gamma v w^4\left(w^2+1-u^2\right)}{2} \nonumber\\ 
&&+\frac{3}{2}w^2(w-v)\left(1+u^2+w^2\right).\label{first-chart-potI-modelII-3}
\end{eqnarray}
As $\frac{du}{dt}=0$,  $\frac{dv}{dt}=-\frac{3}{2}v^3\left(1+u^2\right)$ and 
$\frac{dw}{dt}=0$ at $w=0$, the critical points at infinity are given by 
$\left\{(u,0,0):u\in\mathbb{R}\right\}$, corresponding to the circle 
$\mathbb{S}^1$ with $y_1>0$. Taking into account the phase space domain, the 
critical point at infinity is the point 
$B^{+\infty}\left(\frac{1}{\sqrt{2}},\frac{1}{\sqrt{2}},0,0\right)$, implying 
the point $(1,0,0)$ in the $(u,v,w)$ coordinate system. Linear stability 
analysis fails to predict the stability of the point $(1,0,0)$, since all   
eigenvalues at this point are zero. Nevertheless, at $v=0$ plane we have
$\frac{du}{d\tau}=\frac{1}{2}uw^3\left(6w+\sqrt{6}
\lambda\left(u^2-1\right)\right)$ and 
$\frac{dw}{d\tau}=\frac{1}{2}w^3\left(3\left(1+w^2\right)+u^2(3+\sqrt{6}\lambda 
w)\right)$. Hence, close to $w=0$, we have $\frac{dw}{d\tau}\cong 
\frac{3}{2}w^3\left(1+u^2\right)$, which states that $w$ is increasing for 
$w>0$ and decreasing for $w<0$. As a result,   point $(1,0,0)$ exhibits 
unstable behavior and the corresponding point $B^{+\infty}$ is also unstable. 

In the second chart  we   consider
$\phi_2\left(y_1,y_2,y_3,y_4\right)=(u,v,w)$, leading to the system
\begin{eqnarray}
\frac{du}{d\tau}&=&\frac{3}{2} u^2 w(w-v-1)\left(w^2-u^2-1\right)-3u^2w^3(w-v)  
\nonumber \\ && +\frac{\sqrt{6}}{2}\lambda u w^2\left(u^2-1\right)-\frac{\gamma v w^3\left(w^2+u^2-1\right)}{2},\label{second-chart-potI-modelII-1}\\
\frac{dv}{d\tau} &=&\frac{3}{2}uv(1-w)(w-v)\left(w^2-u^2-1\right) \nonumber \\&& +\frac{\sqrt{6}}{2}\lambda u^2vw^2(w-v),\\
\label{second-chart-potI-modelII-2}
\frac{dw}{d\tau} &=& \frac{\sqrt{6}}{2}\lambda u^2 w^3-\frac{3}{2}u w^2(w-v)\left(w^2-u^2-1\right).\label{second-chart-potI-modelII-3}
\end{eqnarray}
Similarly, the points at infinity are  described by the line 
$\left\{(u,0,0):u\in\mathbb{R}\right\}$, representing the circle 
$\mathbb{S}^{1}$ with $y_1>0$. Thus, the critical point at infinity is 
$\left(\frac{1}{\sqrt{2}},\frac{1}{\sqrt{2}},0,0\right)$, which already has 
been  obtained in the first chart.

In the third chart we    consider the local map 
$\phi_3\left(y_1,y_2,y_3,y_4\right)=(u,v,w)$. Therefore, the system 
(\ref{aut-sys-potI-modelII-1})-(\ref{aut-sys-potI-modelII-3}) in terms of $u$, 
$v$ and $w$ can be expressed as
\begin{eqnarray}
\frac{du}{d\tau}&=&\frac{3}{2} u^2 (w-1)\left(w^2-u^2-v^2\right)-\frac{\sqrt{6}}{2}\lambda u v^2 w^2\left(w-1\right) 
\nonumber \\ && -3u^2w^3(w-1)-\frac{\gamma w^3\left(w^2+u^2-v^2\right)}{2},\label{third-chart-potI-modelII-1}\\
\frac{dv}{d\tau} &=&\frac{3}{2}uv(w-1)\left(w^2-u^2-v^2\right) \nonumber \\&& -\frac{\sqrt{6}}{2}\lambda u^2vw^2(w-1),\\
\label{third-chart-potI-modelII-2}
\frac{dw}{d\tau} &=& -\frac{3}{2}uw(w-1)^2\left(w^2-u^2-v^2\right).\label{third-chart-potI-modelII-3}
\end{eqnarray}
As we can see, for this scenario, no critical points exist at infinity.

In the fourth chart, if we consider 
$\phi_4\left(y_1,y_2,y_3,y_4\right)=(u,v,w)$, the dynamical system becomes: 
\begin{eqnarray}
    \frac{du}{d\tau}=P(u,v,w),\label{fourth-chart-potI-modelII-1}\\
    \frac{dv}{d\tau}=Q(u,v,w),\label{fourth-chart-potI-modelII-2}\\
    \frac{dw}{d\tau}=R(u,v,w).\label{fourth-chart-potI-modelII-3}
\end{eqnarray}
This is the same as the system 
(\ref{reg-aut-sys-potI-modelII-1})-(\ref{reg-aut-sys-potI-modelII-3}), and 
it does not yield any new solutions at infinity.

When working with the phase-space  domain $\mathbb{D}$ where $x<0$, we 
introduce a re-parametrization, defined by $dN=-x(1-z)d\tau$, which 
transforms the system 
(\ref{aut-sys-potI-modelII-1})-(\ref{aut-sys-potI-modelII-3}) as  
 \begin{eqnarray}
\frac{dx}{d\tau}&=&3x^2(1-z)-\frac{3}{2}x^2\left(1-x^2-y^2\right)(1-z)  
\nonumber \\ && +\frac{\sqrt{6}}{2}\lambda x y^2(1-z)+\frac{\gamma z 
\left(1+x^2-y^2\right)}{2},\label{n-reg-aut-sys-potI-modelII-1}\\
\frac{dy}{d\tau} &=& \left[\frac{\sqrt{6}}{2}\lambda x^2 y-\frac{3}{2} x y\left(1-x^2- y^2\right)\right](1-z),\\ 
\label{n-reg-aut-sys-potII-modelII-2}
\frac{dz}{d\tau}&=& -\frac{3}{2} x z(1-z)^2\left(1-x^2-y^2\right).\label{n-reg-aut-sys-potI-modelII-3}
\end{eqnarray}
Following the same approach as in the previous paragraphs, we observe that there exists a
critical point at infinity, namely
$B^{-\infty}\left(-\frac{1}{\sqrt{2}},\frac{1}{\sqrt{2}},0,0\right)$, which 
exhibits   unstable behavior.

In conclusion, the original system 
(\ref{aut-sys-potI-modelII-1})-(\ref{aut-sys-potI-modelII-3})  has two points at 
infinity, namely
$B^{\pm\infty}\left(\pm\frac{1}{\sqrt{2}},\frac{1}{\sqrt{2}},0,0\right)$, both 
representing unstable solutions, and thus they cannot attract the Universe at 
late times.

\section{Potential $V(\phi)=V_0\cosh{(\kappa\eta\phi)}$}

\subsection{Model I: $Q=\alpha H \rho_m$}\label{appendix-3}

For the phase-space domain $\mathbb{D}_+=\mathbb{D}\setminus\{x\leq 0\}$, the re-parametrization, defined as 
$dN=xd\tau$, transforms the systems (\ref{aut-sys-potII-modelI-1})-(\ref{aut-sys-potII-modelI-3}) into the following:
\begin{eqnarray}
   \frac{dx}{d\tau} &=& -3x^2- \frac{\sqrt{6}}{2}\lambda x y^2+\frac{3}{2}x^2\left(1-x^2-y^2\right) 
\nonumber \\ &&-  
\frac{\alpha}{2}\left(1+x^2-y^2\right),\label{reg-aut-sys-potII-modelI-1}\\
    \frac{dy}{d\tau} &=& -\frac{\sqrt{6}}{2}\lambda x^2 
y+\frac{3}{2}xy\left(1-x^2-y^2\right),\label{reg-aut-sys-potII-modelI-2}\\
    \frac{d\lambda}{d\tau} &=&  
\sqrt{6}\left(\lambda^2-\eta^2\right)x^2.\label{reg-aut-sys-potII-modelI-3}
\end{eqnarray}
The vector field corresponding  to the system described above is represented by 
the components $P$, $Q$ and $R$. These are defined as: $P=-3x^2- 
\frac{\sqrt{6}}{2}\lambda x y^2+\frac{3}{2}x^2\left(1-x^2-y^2\right)-  
\frac{\alpha}{2}\left(1+x^2-y^2\right)$, $Q=-\frac{\sqrt{6}}{2}\lambda x^2 
y+\frac{3}{2}xy\left(1-x^2-y^2\right)$ and 
$R=\sqrt{6}\left(\lambda^2-\eta^2\right)x^2$. We project the 
vector field onto the upper three-sphere $\mathbb{S}^{+}$ via central projection 
which is defined by the following relations for $y_1$, $y_2$, $y_3$ and $y_4$: 
$y_1=\frac{x}{\Delta x}$, $y_2=\frac{y}{\Delta x}$, $y_3=\frac{\lambda}{\Delta 
x}$ and $y_4=\frac{1}{\Delta x}$, where $\Delta x=\sqrt{1+x^2+y^2+\lambda^2}$. 
The stability analysis is then carried out in four distinct local charts, 
$\left(U_i,\phi_i\right)$, labeled with $i=1,2,3,4$, and the sets $U_i$ are 
given by $U_i=\left\{\left(y_1,y_2,y_3,y_4\right)\in\mathbb{S}^3:y_i>0\right\}$. 
The local map $\phi_i:U_i\longrightarrow\mathbb{R}^3$ is given by the relation  
$\phi_i\left(y_1,y_2,y_3,y_4\right)=\left(\frac{y_n}{y_i},\frac{y_m}{y_i},\frac{
y_p}{y_i}\right)$, 
with the indices $n$, $m$, $p$ satisfying $n<m<p$, and are all distinct from 
$i$.

For the first chart, by setting $\phi_1\left(y_1,y_2,y_3,y_4\right)=(u,v,w)$,  
the dynamical system 
(\ref{reg-aut-sys-potII-modelI-1})-(\ref{reg-aut-sys-potII-modelI-3}) transforms 
as
\begin{eqnarray}
\frac{du}{d\tau}&=&3uw^2+\frac{\alpha}{2} u w^2\left(1-u^2+w^2\right)  
\nonumber \\ && +\frac{\sqrt{6}}{2}uv\left(u^2-1\right),\label{first-chart-potII-modelI-1}\\
\frac{dv}{d\tau} &=&\frac{3}{2}v\left(1+u^2+w^2\right)+\frac{\alpha}{2}vw^2\left(1-u^2+w^2\right) \nonumber \\&& +\frac{\sqrt{6}}{2} u^2 v^2+\sqrt{6}\left(v^2-\eta^2 w^2\right),\\
\label{first-chart-potII-modelI-2}
\frac{dw}{d\tau} &=& \frac{3}{2}w\left(1+u^2+w^2\right)+\frac{\alpha}{2}w^3\left(1-u^2+w^2\right) \nonumber\\ 
&&+\frac{\sqrt{6}}{2}u^2 v w.\label{first-chart-potII-modelI-3}
\end{eqnarray}
 At $w=0$ we have $\frac{du}{d\tau}=\frac{\sqrt{6}}{2}uv\left(u^2-1\right)$, 
$\frac{dv}{d\tau}=\frac{3}{2}uv\left(\sqrt{6}uv+1+u^2\right)$ and 
$\frac{dw}{d\tau}=0$, which indicates that the points at infinity of the above 
system are $\left\{(u,0,0):u\in\mathbb{R}\right\}$, 
$\left\{(0,v,0):v\in\mathbb{R}\right\}$, $\left(1,-\sqrt{\frac{2}{3}},0\right)$ 
and $\left(-1,\sqrt{\frac{2}{3}},0\right)$. From the perspective of our 
phase-space domain, the  critical points at infinity which lie on 
$\mathbb{S}^3$ are 
$C_1^{+\infty}\left(\frac{1}{\sqrt{2}},\frac{1}{\sqrt{2}},0,0\right)$, and
$C_2^{+\infty}\left(\sqrt{\frac{3}{8}},\sqrt{\frac{3}{8}},-\frac{1}{2},0
\right)$, indicating respectively the points $(1,0,0)$ and 
$\left(1,-\sqrt{\frac{2}{3}},0\right)$  in the $(u,v,w)$ 
coordinate  system. The eigenvalues at $(1,0,0)$ are $3,3,0$, concluding it is 
unstable in nature. Similarly, the eigenvalues evaluated at 
$\left(1,-\sqrt{\frac{2}{3}},0\right)$ are $-3,-2,2$, implying that this point 
is  saddle. Therefore, $C_1^{+\infty}$ exhibits unstable 
nature, while $C_2^{+\infty}$  exhibits  saddle behavior. 

 In the second chart we find two points at infinity, namely,  
$C_1^{+\infty}\left(\frac{1}{\sqrt{2}},\frac{1}{\sqrt{2}},0,0\right)$, and
$C_2^{+\infty}\left(\sqrt{\frac{3}{8}},\sqrt{\frac{3}{8}},-\frac{1}{2},0
\right)$,
which we have already seen in the first chart. Moreover, we cannot detect any 
points at infinity in the third chart for our phase space domain. Finally, the 
dynamical system in the fourth chart is completely analogous to the system 
(\ref{reg-aut-sys-potII-modelI-1})-(\ref{reg-aut-sys-potII-modelI-3}), which 
does not lead to  any new solutions at infinity.

In the phase-space domain $\mathbb{D}_-=\mathbb{D}\setminus\{x\geq 0\}$, the re-parametrization, defined as 
$dN=-xd\tau$, leads to a simplified version of the system of 
equations (\ref{aut-sys-potII-modelI-1})-(\ref{aut-sys-potI-modelII-3}), namely
\begin{eqnarray}
   \frac{dx}{d\tau} &=& 3x^2+\frac{\sqrt{6}}{2}\lambda x y^2-\frac{3}{2}x^2\left(1-x^2-y^2\right) 
\nonumber \\ &&+  
\frac{\alpha}{2}\left(1+x^2-y^2\right),\label{n-reg-aut-sys-potII-modelI-1}\\
    \frac{dy}{d\tau} &=& \frac{\sqrt{6}}{2}\lambda x^2 
y-\frac{3}{2}xy\left(1-x^2-y^2\right),\label{n-reg-aut-sys-potII-modelI-2}\\
    \frac{d\lambda}{d\tau} &=&  
-\sqrt{6}\left(\lambda^2-\eta^2\right)x^2.\label{n-reg-aut-sys-potII-modelI-3}
\end{eqnarray}
Following the same reasoning as in the previous paragraphs,  we extract two 
critical points at infinity, namely
$C_1^{-\infty}\left(-\frac{1}{\sqrt{2}},\frac{1}{\sqrt{2}},0,0\right)$ and 
$C_2^{-\infty}\left(-\sqrt{\frac{3}{8}},\sqrt{\frac{3}{8}},\frac{1}{2},
0\right)$, and we deduce that $C_1^{-\infty}$ is unstable, while 
$C_2^{-\infty}$ is saddle.

To summarize,  the original system 
(\ref{aut-sys-potII-modelI-1})-(\ref{aut-sys-potII-modelI-3}) has four points at 
infinity, i.e.
$C_1^{\pm\infty}\left(\pm\frac{1}{\sqrt{2}},\frac{1}{\sqrt{2}},0,0\right)$, and
$C_2^{\pm\infty}\left(\pm\sqrt{\frac{3}{8}},\sqrt{\frac{3}{8}},\mp\frac{1}{2},
0\right)$, which exhibit unstable dynamics, and thus they cannot attract the 
Universe at late times.

\subsection{Model II: $Q=\Gamma \rho_m$}\label{appendix-4}

Making use of the re-parametrization, defined as $dN=x(1-z)d\tau$ for the phase-space domain $\mathbb{D}_+=\mathbb{D}\setminus\{x\leq0\}$,  the 
system of equations 
(\ref{aut-sys-potII-modelII-1})-(\ref{aut-sys-potII-modelII-4}) becomes  
\begin{eqnarray}
   \frac{dx}{d\tau}&=&-3x^2(1-z)+\frac{3}{2}x^2\left(1-x^2-y^2\right)(1-z)  
\nonumber \\ && -\frac{\sqrt{6}}{2}\lambda x y^2(1-z)-\frac{\gamma z 
\left(1+x^2-y^2\right)}{2},\label{reg-aut-sys-potII-modelII-1}\\
    \frac{dy}{d\tau} &=& \left[-\frac{\sqrt{6}}{2}\lambda x^2 
y+\frac{3}{2}xy\left(1-x^2-y^2\right)\right](1-z),\label{reg-aut-sys-potII-modelII-2}\\
    \frac{d\lambda}{d\tau} &=&  
\sqrt{6}\left(\lambda^2-\eta^2\right)x^2(1-z),\label{reg-aut-sys-potII-modelII-3}\\
\frac{dz}{d\tau} &=& \frac{3}{2} x z(1-z)^2\left(1-x^2-y^2\right). 
\label{reg-aut-sys-potII-modelII-4}
\end{eqnarray}
The vector field corresponding to the system is represented  by the components 
$P$, $Q$, $R$ and $S$,   defined as: 
$P=-3x^2(1-z)+\frac{3}{2}x^2\left(1-x^2-y^2\right)(1-z)-\frac{\sqrt{6}}{2}
\lambda x y^2(1-z)-\frac{\gamma z 
\left(1+x^2-y^2\right)}{2}$, $Q=\left[-\frac{\sqrt{6}}{2}\lambda x^2 
y+\frac{3}{2}xy\left(1-x^2-y^2\right)\right](1-z)$,  
$R=\sqrt{6}\left(\lambda^2-\eta^2\right)x^2(1-z)$ and $S=\frac{3}{2} x 
z(1-z)^2\left(1-x^2-y^2\right)$. We start by projecting the vector field onto 
the upper four dimensional sphere $\mathbb{S}^{4+}$ using central projection, defined by the following relations for $y_1$, $y_2$, $y_3$, $y_4$, $y_5$: 
$y_1=\frac{x}{\Delta x}$, $y_2=\frac{y}{\Delta x}$, $y_3=\frac{\lambda}{\Delta 
x}$, $y_4=\frac{z}{\Delta x}$ and $y_5=\frac{1}{\Delta x}$, where $\Delta 
x=\sqrt{1+x^2+y^2+\lambda^2+z^2}$. The stability of the system is studied within 
five distinct local charts $\left(U_i,\phi_i\right)$ indexed by $i=1,2,3,4,5$, 
where the sets $U_i$ are defined as
$U_i=\left\{\left(y_1,y_2,y_3,y_4,y_5\right)\in\mathbb{S}^4:y_i>0\right\}$. The 
local map $\phi_i:U_i\longrightarrow\mathbb{R}^4$ is given by 
$\phi_i\left(y_1,y_2,y_3,y_4,y_5\right)=\left(\frac{y_n}{y_i},\frac{y_m}{y_i},
\frac{y_p}{y_i},\frac{y_q}{y_i}\right)$, where $n$, $m$, $p$ and $q$ are 
distinct indices such that $n<m<p<q$,  and none of $n, m, p, q$ are equal to 
$i$.

For the  first chart, by selecting 
$\phi_1\left(y_1,y_2,y_3,y_4,y_5\right)=(u,v,w,r)$, the dynamical system 
described by equations 
(\ref{reg-aut-sys-potII-modelII-1})-(\ref{reg-aut-sys-potII-modelII-4}) is 
modified as  
\begin{eqnarray}
\frac{du}{d\tau}&=&3ur^3(r-w)+\frac{\gamma u w r^3(1-u^2+r^2)}{2}  
\nonumber \\ && +\frac{\sqrt{6}}{2} u v r\left(u^2-1\right)(r-w),\label{first-chart-potII-modelII-1}\\
\frac{dv}{d\tau} &=&\frac{3}{2}vr(r-w)\left(1+u^2+r^2\right)  \nonumber \\&& 
+\frac{\sqrt{6}}{2}u^2v^2r(r-w)+\frac{\gamma v w r^3\left(1-u^2+r^2\right)}{2} 
\nonumber \\
&&+\sqrt{6}r(r-w)\left(v^2-\eta^2r^2\right),\\
\label{first-chart-potII-modelII-2}
\frac{dw}{d\tau} &=&\frac{3}{2}wr(r-w)\left(1+u^2+r^2\right) \nonumber \\&& +\frac{\sqrt{6}}{2}u^2vwr(r-w)+\frac{\gamma w^2 r^3\left(1-u^2+r^2\right)}{2} \nonumber \\
&&+\frac{3}{2}w(r-w)^2\left(r^2-1-u^2\right),\\
\label{first-chart-potII-modelII-3}
\frac{dr}{d\tau} &=& \frac{3}{2}r^2(r-w)\left(1+u^2+r^2\right)+\frac{\sqrt{6}}{2}u^2vr^2(r-w) \nonumber\\ 
&&+\frac{\gamma w r^4\left(1-u^2+r^2\right)}{2}.\label{first-chart-potII-modelII-4}
\end{eqnarray}
At $r=0$, the  system evolves as   $\frac{du}{d\tau}=0$, 
$\frac{dv}{d\tau}=0$, $\frac{dw}{d\tau}=-\frac{3}{2}w^3\left(1+u^2\right)$ and 
$\frac{dr}{d\tau}=0$. This indicates that the critical points at infinity 
are given by $\left\{(u,v,0,0):u\in\mathbb{R},v\in\mathbb{R}\right\}$. From the 
perspective of our phase-space domain, the critical points at infinity, located 
on the sphere $\mathbb{S}^4$, are 
$D^{+\infty}\Bigl\{\left(y_1,y_2,y_3,y_4,y_5\right)\in\mathbb{S}^4:y_5=0,y_4=0,
y_1^2=y_2^2, y_1>0,y_2>0\Bigr\}$. At the points $(u,v,0,0)$ all   eigenvalues 
are zero, which suggests that a linear stability analysis does not provide 
conclusive information about stability. However, on the $w=0$ surface  the 
system evolves as   
$\frac{du}{d\tau}=3r^4u+\sqrt{\frac{3}{2}}r^2uv\left(u^2-1\right)$, 
$\frac{dv}{d\tau}=\frac{1}{2}r^2\left(3v\left(1+r^2+u^2\right)+\sqrt{6}
v^2\left(2+u^2\right)-2\sqrt{6}r^2\eta^2\right)$, $\frac{dw}{d\tau}=0$ and 
$\frac{dr}{d\tau}=\frac{1}{2}r^3\left(3+3r^2+u^2(3+\sqrt{6}v)\right)$. From 
these  we observe that $r$ increases when $r>0$ with $v>0$, and decreases when 
$r<0$ with $v>0$. As a result, $D^{+\infty}$ exhibits unstable behavior.

The second  and third charts reveal the same point at infinity, namely  
$D^{+\infty}$, as found in the first chart. Furthermore, the fourth 
chart does not uncover any additional critical points at infinity within our phase space 
domain. Finally, the fifth chart yields a dynamical system similar to 
(\ref{reg-aut-sys-potII-modelII-1})-(\ref{reg-aut-sys-potII-modelII-4}), without 
generating any new solutions at infinity.

By using the re-parametrization $dN=-x(1-z)d\tau$ in the phase-space domain 
$\mathbb{D}_-=\mathbb{D}\setminus\{x\geq 0\}$, we obtain the following simplified form of the system 
(\ref{aut-sys-potII-modelII-1})-(\ref{aut-sys-potII-modelII-4}), namely, 
\begin{eqnarray}
   \frac{dx}{d\tau}&=&3x^2(1-z)-\frac{3}{2}x^2\left(1-x^2-y^2\right)(1-z)  
\nonumber \\ && +\frac{\sqrt{6}}{2}\lambda x y^2(1-z)+\frac{\gamma z 
\left(1+x^2-y^2\right)}{2},\label{n-reg-aut-sys-potII-modelII-1}\\
    \frac{dy}{d\tau} &=& \left[\frac{\sqrt{6}}{2}\lambda x^2 
y-\frac{3}{2}xy\left(1-x^2-y^2\right)\right](1-z),\label{n-reg-aut-sys-potII-modelII-2}\\
    \frac{d\lambda}{d\tau} &=&  
-\sqrt{6}\left(\lambda^2-\eta^2\right)x^2(1-z),\label{n-reg-aut-sys-potII-modelII-3}\\
\frac{dz}{d\tau} &=& -\frac{3}{2} x z(1-z)^2\left(1-x^2-y^2\right).\label{n-reg-aut-sys-potII-modelII-4}
\end{eqnarray}
Continuing  with the same approach as in previous sections, we identify the 
critical points at infinity as  
$D^{-\infty}\Bigl\{\left(y_1,y_2,y_3,y_4,y_5\right)\in\mathbb{S}^4:y_5=0,y_4=0,
y_1^2=y_2^2,\\ y_1<0,y_2>0\Bigr\}$. This suggests that $D^{-\infty}$ also 
exhibits unstable characteristics.

In conclusion,  the original system 
(\ref{aut-sys-potII-modelII-1})-(\ref{aut-sys-potII-modelII-4}) has two points 
at infinity: $D^{+\infty}$ and $D^{-\infty}$, both of which exhibit unstable 
dynamics, and thus they cannot be the late-time solutions of the Universe.

\section{Potential $V (\phi) = \frac{1}{8 \beta} \left(1 - e^{- \kappa 
\mu\; 
\phi}\right)^2$}

\subsection{Model I: $Q=\alpha H \rho_m$}\label{appendix-5}

For the system of 
equations (\ref{aut-sys-potIII-modelI-1})-(\ref{aut-sys-potIII-modelI-3}), the re-parametrization $dN=xd\tau$ in the
phase-space domain $\mathbb{D}_+=\mathbb{D}\setminus\{x\leq0\}$  yields the following:
\begin{eqnarray}
   \frac{dx}{d\tau} &=& -3x^2- \frac{\sqrt{6}}{2}\lambda x y^2+\frac{3}{2}x^2\left(1-x^2-y^2\right) 
\nonumber \\ &&-  
\frac{\alpha}{2}\left(1+x^2-y^2\right),\label{reg-aut-sys-potIII-modelI-1}\\
    \frac{dy}{d\tau} &=& -\frac{\sqrt{6}}{2}\lambda x^2 
y+\frac{3}{2}xy\left(1-x^2-y^2\right),\label{reg-aut-sys-potIII-modelI-2}\\
    \frac{d\lambda}{d\tau} &=&  
\frac{\sqrt{6}}{2}\left(\lambda^2-2\mu\lambda\right)x^2.\label{reg-aut-sys-potIII-modelI-3}
\end{eqnarray}
The components  $P$, $Q$ and $R$ represent the vector field corresponding to the 
system outlined above. These are given by  $P=-3x^2- \frac{\sqrt{6}}{2}\lambda 
x y^2+\frac{3}{2}x^2\left(1-x^2-y^2\right)-  
\frac{\alpha}{2}\left(1+x^2-y^2\right)$, $Q=-\frac{\sqrt{6}}{2}\lambda x^2 
y+\frac{3}{2}xy\left(1-x^2-y^2\right)$ and 
$R=\frac{\sqrt{6}}{2}\left(\lambda^2-2\mu\lambda\right)x^2$.  We begin by 
projecting the vector field onto the upper three dimensional sphere $\mathbb{S}^{3+}$ using 
central projection, defined by the following relations for $y_1$, $y_2$, 
$y_3$ and $y_4$: $y_1=\frac{x}{\Delta x}$, $y_2=\frac{y}{\Delta x}$, 
$y_3=\frac{\lambda}{\Delta x}$ and $y_4=\frac{1}{\Delta x}$. The stability is 
then analyzed within four distinct local charts $\left(U_i,\phi_i\right)$ with 
indices $i=1,2,3,4$, where the sets $U_i$ are given by  
$U_i=\left\{\left(y_1,y_2,y_3,y_4\right)\in\mathbb{S}^3:y_i>0\right\}$. The 
local map $\phi_i:U_i\longrightarrow\mathbb{R}^3$ is described by 
$\phi_i\left(y_1,y_2,y_3,y_4\right)=\left(\frac{y_n}{y_i},\frac{y_m}{y_i},\frac{
y_p}{y_i}\right)$, with the indices $n$, $m$ and $p$ satisfying $n<m<p$, with 
$n, m, p\neq i$.

In the case of the first chart,  by choosing 
$\phi_1\left(y_1,y_2,y_3,y_4\right)=(u,v,w)$  the dynamical system 
(\ref{reg-aut-sys-potIII-modelI-1})-(\ref{reg-aut-sys-potIII-modelI-3}) becomes
\begin{eqnarray}
\frac{du}{d\tau}&=&3uw^2+\frac{\alpha}{2} u w^2\left(1-u^2+w^2\right)  
\nonumber \\ && +\frac{\sqrt{6}}{2}uv\left(u^2-1\right),\label{first-chart-potIII-modelI-1}\\
\frac{dv}{d\tau} &=&\frac{3}{2}v\left(1+u^2+w^2\right) 
+\frac{\alpha}{2}vw^2\left(1-u^2+w^2\right) \nonumber \\&& +\frac{\sqrt{6}}{2} 
u^2 v^2+\frac{\sqrt{6}}{2}\left(v^2-2\mu vw\right),\\
\label{first-chart-potIII-modelI-2}
\frac{dw}{d\tau} &=& 
\frac{3}{2}w\left(1+u^2+w^2\right)+\frac{\alpha}{2}w^3\left(1-u^2+w^2\right)  
\nonumber\\ 
&&+\frac{\sqrt{6}}{2}u^2 v w.\label{first-chart-potIII-modelI-3}
\end{eqnarray}
At $w=0$, the system evolves  according to 
$\frac{du}{d\tau}=\frac{\sqrt{6}}{2}uv\left(u^2-1\right)$, 
$\frac{dv}{d\tau}=\frac{1}{2}v\left(1+u^2\right)(\sqrt{6}v+3)$ and 
$\frac{dw}{d\tau}=0$. This suggests that the critical points at infinity  
are $\left\{(u,0,0):u\in\mathbb{R}\right\}$, 
$\left(0,-\sqrt{\frac{3}{2}},0\right)$, $\left(1,-\sqrt{\frac{3}{2}},0\right)$ 
and $\left(-1,-\sqrt{\frac{3}{2}},0\right)$. From the viewpoint of our phase 
space domain, the critical points at infinity located on $\mathbb{S}^3$ are 
$E_1^{+\infty}\left(\frac{1}{\sqrt{2}},\frac{1}{\sqrt{2}},0,0\right)$   and
$E_2^{+\infty}\left(\sqrt{\frac{2}{7}},\sqrt{\frac{2}{7}},-\sqrt{\frac{3}{7}},
0\right)$, corresponding respectively to the points $(1,0,0)$ and 
$\left(1,-\sqrt{\frac{3}{2}},0\right)$   in   $(u,v,w)$ 
coordinates. The eigenvalues for the point $(1,0,0)$ are $3,3$ and $0$, which 
indicates an unstable nature. Additionally, at   point 
$\left(1,-\sqrt{\frac{3}{2}},0\right)$ the eigenvalues are $-3,-3$ and 
$\frac{3}{2}$, suggesting that it is a saddle point. Consequently, 
$E_1^{+\infty}$ is unstable, and $E_2^{+\infty}$ behaves as a saddle point.

In the  second chart  we again find two points at infinity, namely
$E_1^{+\infty}\left(\frac{1}{\sqrt{2}},\frac{1}{\sqrt{2}},0,0\right)$ and 
$E_2^{+\infty}\left(\sqrt{\frac{2}{7}},\sqrt{\frac{2}{7}},-\sqrt{\frac{3}{7}},
0\right)$, which were identified in the first chart. The third chart does not 
reveal any additional points at infinity in our phase-space domain. The fourth 
chart gives a dynamical system that is similar to the system 
(\ref{reg-aut-sys-potIII-modelI-1})-(\ref{reg-aut-sys-potIII-modelI-3}), and 
therefore it does not provide any new solutions at infinity.

In  the phase space domain 
$\mathbb{D}_-=\mathbb{D}\setminus\{x\geq 0\}$, we employ the re-parametrization 
$dN=-xd\tau$ and obtain a simplified version of the system 
(\ref{aut-sys-potIII-modelI-1})-(\ref{aut-sys-potIII-modelI-3}) as
\begin{eqnarray}
   \frac{dx}{d\tau} &=& 3x^2+\frac{\sqrt{6}}{2}\lambda x y^2-\frac{3}{2}x^2\left(1-x^2-y^2\right) 
\nonumber \\ &&+  
\frac{\alpha}{2}\left(1+x^2-y^2\right),\label{n-reg-aut-sys-potIII-modelI-1}\\
    \frac{dy}{d\tau} &=& \frac{\sqrt{6}}{2}\lambda x^2 
y-\frac{3}{2}xy\left(1-x^2-y^2\right),\label{n-reg-aut-sys-potIII-modelI-2}\\
    \frac{d\lambda}{d\tau} &=&  
-\frac{\sqrt{6}}{2}\left(\lambda^2-2\mu\lambda\right)x^2.\label{n-reg-aut-sys-potIII-modelI-3}
\end{eqnarray}
Continuing with the same approach as in previous sections,  we find two critical 
points at infinity: 
$E_1^{-\infty}\left(-\frac{1}{\sqrt{2}},\frac{1}{\sqrt{2}},0,0\right)$ and 
$E_2^{-\infty}\left(-\sqrt{\frac{2}{7}},\sqrt{\frac{2}{7}},\sqrt{\frac{3}{7}},
0\right)$. This suggests that $E_1^{-\infty}$ is unstable, while $E_2^{-\infty}$ 
is saddle.

To conclude, the original system  
(\ref{aut-sys-potIII-modelI-1})-(\ref{aut-sys-potIII-modelI-3}) has four points 
at infinity: 
$E_1^{\pm\infty}\left(\pm\frac{1}{\sqrt{2}},\frac{1}{\sqrt{2}},0,0\right)$, and
$E_2^{\pm\infty}\left(\pm\sqrt{\frac{2}{7}},\sqrt{\frac{2}{7}},\mp\sqrt{\frac{3}
{7}},0\right)$, all of which display unstable dynamics. Therefore, they cannot 
be the late-time solutions of the Universe.

\subsection{Model II: $Q=\Gamma \rho_m$}\label{appendix-6}

After using a re-parametrization of the form $dN=x(1-z)d\tau$ in
the phase-space domain $\mathbb{D}_+=\mathbb{D}\setminus\{x\leq0\}$ for the system of equations
 (\ref{aut-sys-potIII-modelII-1})-(\ref{aut-sys-potIII-modelII-4}), we get
\begin{eqnarray}
\frac{dx}{d\tau}&=&-3x^2(1-z)+\frac{3}{2}x^2\left(1-x^2-y^2\right)(1-z)  
\nonumber \\ && -\frac{\sqrt{6}}{2}\lambda x y^2(1-z)-\frac{\gamma z 
\left(1+x^2-y^2\right)}{2},\label{reg-aut-sys-potIII-modelII-1}\\
    \frac{dy}{d\tau} &=& \left[-\frac{\sqrt{6}}{2}\lambda x^2 
y+\frac{3}{2}xy\left(1-x^2-y^2\right)\right](1-z),\label{reg-aut-sys-potIII-modelII-2}\\
    \frac{d\lambda}{d\tau} &=&  
\frac{\sqrt{6}}{2}\left(\lambda^2-2\mu\lambda\right)x^2(1-z),\label{reg-aut-sys-potIII-modelII-3}\\
\frac{dz}{d\tau} &=&  \frac{3}{2} x 
z(1-z)^2\left(1-x^2-y^2\right).\label{reg-aut-sys-potIII-modelII-4}
\end{eqnarray}
The components $P$, $Q$, $R$ and $S$ represent the vector field  of the above 
system, and are expressed as  
$P=-3x^2(1-z)+\frac{3}{2}x^2\left(1-x^2-y^2\right)(1-z)-\frac{\sqrt{6}}{2}
\lambda x y^2(1-z)-\frac{\gamma z 
\left(1+x^2-y^2\right)}{2}$, $Q=\left[-\frac{\sqrt{6}}{2}\lambda x^2 
y+\frac{3}{2}xy\left(1-x^2-y^2\right)\right](1-z)$,  
$R=\frac{\sqrt{6}}{2}\left(\lambda^2-2\mu\lambda\right)x^2(1-z)$ and 
$S=\frac{3}{2} x z(1-z)^2\left(1-x^2-y^2\right)$. We first project the vector 
field onto the upper four dimensional sphere $\mathbb{S}^{4+}$ via central projection, where the relations for $y_1$, $y_2$, $y_3$, $y_4$ and $y_5$ are given by: 
$y_1=\frac{x}{\Delta x}$, $y_2=\frac{y}{\Delta x}$, $y_3=\frac{\lambda}{\Delta 
x}$, $y_4=\frac{z}{\Delta x}$ and $y_5=\frac{1}{\Delta x}$, where $\Delta 
x=\sqrt{1+x^2+y^2+\lambda^2+z^2}$. To analyze the stability, we use five 
distinct local charts $\left(U_i,\phi_i\right)$, where $i=1,2,3,4,5$ and the 
sets $U_i$ are defined as 
$U_i=\left\{\left(y_1,y_2,y_3,y_4,y_5\right)\in\mathbb{S}^4:y_i>0\right\}$. The 
local map $\phi_i:U_i\longrightarrow\mathbb{R}^4$ is specified by 
$\phi_i\left(y_1,y_2,y_3,y_4,y_5\right)=\left(\frac{y_n}{y_i},\frac{y_m}{y_i},
\frac{y_p}{y_i},\frac{y_q}{y_i}\right)$, where the indices $n$, $m$, $p$ and $q$ 
satisfy $n<m<p<q$, while $n, m, p, q\neq i$.

In the first chart,  by defining 
$\phi_1\left(y_1,y_2,y_3,y_4,y_5\right)=(u,v,w,r)$, the dynamical system  
(\ref{reg-aut-sys-potIII-modelII-1})-(\ref{reg-aut-sys-potIII-modelII-4}) takes 
the   form 
\begin{eqnarray}
\frac{du}{d\tau}&=&\frac{1}{2}ru\Bigl[6r^3+\sqrt{6}rv\left(u^2-1\right)-\sqrt{6}vw\left(u^2-1\right) \nonumber \\ &&+\gamma r^4w+r^2w\left(\gamma-6-\gamma u^2\right)\Bigr],\label{first-chart-potIII-modelII-1}\\
\frac{dv}{d\tau} &=&\frac{1}{2}rv\Bigl[3r^3+(r-w)\left(3+u^2(3+\sqrt{6}v)\right) \nonumber \\ &&+\gamma w r^4-wr^2\left(3+\gamma\left(u^2-1\right)\right) \nonumber \\ &&+\sqrt{6}(r-w)(v-2\mu r)\Bigr],\\
\label{first-chart-potIII-modelII-2}
\frac{dw}{d\tau} &=&\frac{1}{2}w\Bigl[6r^4+wr\left(3+u^2(3-\sqrt{6}v)\right) \nonumber \\ &&-3w^2\left(1+u^2\right)+r^2\left(3w^2+\sqrt{6}u^2v\right) \nonumber \\ &&+\gamma w r^5+wr^3\left(\gamma-9-\gamma u^2\right)\Bigr],\\
\label{first-chart-potIII-modelII-3}
\frac{dr}{d\tau} &=& \frac{1}{2}r^2\Bigl[3r^3+ 
(r-w)\left(3+u^2(3+\sqrt{6}v)\right) \nonumber \\ &&+\gamma w 
r^4+wr^2\left(\gamma-3-\gamma 
u^2\right)\Bigr].\label{first-chart-potIII-modelII-4}
\end{eqnarray}
When $r=0$, the system evolves according to the   set of equations  
$\frac{du}{d\tau}=0$, $\frac{dv}{d\tau}=0$, 
$\frac{dw}{d\tau}=-\frac{3}{2}w^3\left(1+u^2\right)$ and $\frac{dr}{d\tau}=0$. 
This suggests that the points at infinity  are 
$\left\{(u,v,0,0):u\in\mathbb{R},v\in\mathbb{R}\right\}$. Regarding our 
phase-space domain, the critical points at infinity on the four-sphere 
$\mathbb{S}^4$ are given by 
$F^{+\infty}\Bigl\{\left(y_1,y_2,y_3,y_4,y_5\right)\in\mathbb{S}^4:y_5=0,y_4=0, 
y_1^2=y_2^2, y_1>0,y_2>0\Bigr\}$. At the points $(u,v,0,0)$  all eigenvalues 
are zero, indicating that linear stability analysis is inconclusive. However, on 
the surface $w=0$, the system evolves according to the   equations 
$\frac{du}{d\tau}=3r^4u+\sqrt{\frac{3}{2}}r^2uv\left(u^2-1\right)$, 
$\frac{dv}{d\tau}=\frac{1}{2}r^2v\left(3r^2+\left(1+u^2\right)(3+\sqrt{6} 
v)-2\sqrt{6}\mu r\right)$, $\frac{dw}{d\tau}=0$ and 
$\frac{dr}{d\tau}=\frac{1}{2}r^3\left(3+3r^2+u^2(3+\sqrt{6}v)\right)$. This 
leads to the conclusion that $r$   increases for $r>0$ and $v>0$, and decreases 
for $r<0$ and $v>0$. Consequently, $F^{+\infty}$ demonstrates unstable behavior.

In the second and third charts we identify 
$F^{+\infty}$ as the critical point at infinity for our phase space domain, which was already observed in the first chart. The fourth 
chart does not present any new points at infinity. 
The fifth chart leads to a dynamical system that resembles system 
(\ref{reg-aut-sys-potIII-modelII-1})-(\ref{reg-aut-sys-potIII-modelII-4}), 
offering no new solutions at infinity, too.

By using the re-parametrization $dN=-x(1-z)d\tau$ in the phase-space domain 
$\mathbb{D}_-=\mathbb{D}\setminus\{x\geq 0\}$ through 
$dN=-x(1-z)d\tau$, we write the system  
(\ref{aut-sys-potIII-modelII-1})-(\ref{aut-sys-potIII-modelII-4}) as 
\begin{eqnarray}
   \frac{dx}{d\tau}&=&3x^2(1-z)-\frac{3}{2}x^2\left(1-x^2-y^2\right)(1-z)  
\nonumber \\ && +\frac{\sqrt{6}}{2}\lambda x y^2(1-z)+\frac{\gamma z 
\left(1+x^2-y^2\right)}{2},\label{n-reg-aut-sys-potIII-modelII-1}\\
    \frac{dy}{d\tau} &=& \left[\frac{\sqrt{6}}{2}\lambda x^2 
y-\frac{3}{2}xy\left(1-x^2-y^2\right)\right](1-z),\label{n-reg-aut-sys-potIII-modelII-2}\\
    \frac{d\lambda}{d\tau} &=&  
-\frac{\sqrt{6}}{2}\left(\lambda^2-2\mu\lambda\right)x^2(1-z),\label{n-reg-aut-sys-potIII-modelII-3}\\
\frac{dz}{d\tau} &=& -\frac{3}{2} x z(1-z)^2\left(1-x^2-y^2\right).\label{n-reg-aut-sys-potIII-modelII-4}
\end{eqnarray}
Following the  same method as in the previous sections, we determine the 
critical points at infinity as  
$F^{-\infty}\Bigl\{\left(y_1,y_2,y_3,y_4,y_5\right)\in\mathbb{S}^4:y_5=0,y_4=0,
y_1^2=y_2^2,\\ y_1<0,y_2>0\Bigr\}$. This indicates that $F^{-\infty}$  is  
unstable.

In summary, the original system 
(\ref{aut-sys-potIII-modelII-1})-(\ref{aut-sys-potIII-modelII-4}) exhibits two 
points at infinity, namelt $F^{+\infty}$ and $F^{-\infty}$, each of which 
demonstrates unstable behavior, and thus cannot be the late-time stages of the 
Universe.

\section{Potential $V(\phi)=c_0+c_1 e^{\sqrt{\frac{2\kappa^2}{3}}\phi}+c_2 
e^{2\sqrt{\frac{2\kappa^2}{3}}\phi}$}

\subsection{Model I: $Q=\alpha H \rho_m$}\label{appendix-7}

By introducing the re-parametrization $dN=xd\tau$ in the phase space 
domain $\mathbb{D}_+=\mathbb{D}\setminus\{x\leq0\}$, the system  
(\ref{aut-sys-potIV-modelI-1})-(\ref{aut-sys-potIV-modelI-4})  becomes
\begin{eqnarray}
    \frac{dx}{d\tau}&=&-\frac{3}{2}x^2 \left(1+x^2+v^2+w^2+ 
\xi^2\right)+xw^2+2x\xi^2 \nonumber \\ && 
-\frac{\alpha\left(1+x^2-v^2-w^2-\xi^2\right)}{2},
\label{reg-aut-sys-potIV-modelI-1}  \\
\frac{dv}{d\tau}&=&\frac{3}{2}xv\left(1-x^2-v^2-w^2-\xi^2\right),\label{reg-aut-sys-potIV-modelI-2}
\\ 
  \frac{dw}{d\tau}&=&xw\left[x+\frac{3}{2}\left(1-x^2-v^2-w^2-\xi^2\right)\right],
  \label{reg-aut-sys-potIV-modelI-3}\\  
\frac{d\xi}{d\tau}&=&x\xi\left[2x+\frac{3}{2}\left(1-x^2-v^2-w^2-\xi^2\right)\right].
\label{reg-aut-sys-potIV-modelI-4}
\end{eqnarray}
The system  outlined above is described by the vector field components $P$, $Q$, 
$R$ and $S$,  which are given by: $P=-\frac{3}{2}x^2 
\left(1+x^2+v^2+w^2+\xi^2\right)+xw^2+2x\xi^2-\frac{
\alpha\left(1+x^2-v^2-w^2-\xi^2\right)}{2}$, 
$Q=\frac{3}{2}xv\left(1-x^2-v^2-w^2-\xi^2\right)$, 
$R=xw\left[x+\frac{3}{2}\left(1-x^2-v^2-w^2-\xi^2\right)\right]$ and 
$S=x\xi\left[2x+\frac{3}{2}\left(1-x^2-v^2-w^2-\xi^2\right)\right]$. The 
projection of the vector field onto the upper four dimensional sphere $\mathbb{S}^{4+}$ is 
performed using central projection, with the following relations defining 
$y_1$, $y_2$, $y_3$, $y_4$ and $y_5$: $y_1=\frac{x}{\Delta x}$, 
$y_2=\frac{v}{\Delta x}$, $y_3=\frac{w}{\Delta x}$, $y_4=\frac{\xi}{\Delta x}$ 
and $y_5=\frac{1}{\Delta x}$, where $\Delta x=\sqrt{1+x^2+v^2+w^2+\xi^2}$. For 
the stability analysis  we consider five distinct local charts 
$\left(U_i,\phi_i\right)$, indexed by $i=1,2,3,4,5$, with the sets $U_i$ defined 
as  $U_i=\left\{\left(y_1,y_2,y_3,y_4,y_5\right)\in\mathbb{S}^4:y_i>0\right\}$. 
The corresponding local map $\phi_i:U_i\longrightarrow\mathbb{R}^4$ is described 
by 
$\phi_i\left(y_1,y_2,y_3,y_4,y_5\right)=\left(\frac{y_n}{y_i},\frac{y_m}{y_i},
\frac{y_p}{y_i},\frac{y_q}{y_i}\right)$, where the indices $n$, $m$, $p$, $q$ 
follow the condition $n<m<p<q$, and no index among $n, m, p, q$ equals $i$.

By choosing  
$\phi_1\left(y_1,y_2,y_3,y_4,y_5\right)=\left(x_1,x_2,x_3,x_4\right)$ for the 
first chart, the dynamical system 
(\ref{reg-aut-sys-potIV-modelI-1})-(\ref{reg-aut-sys-potIV-modelI-4}) is 
simplified into  
\begin{eqnarray}
\frac{dx_1}{d\tau}&=&\frac{1}{2}x_1x_4 
\Bigl[6x_4-2\left(x_2^2+2x_3^2\right)+\alpha x_4^3 \nonumber \\ &&+\alpha 
x_4\left(1-x_1^2-x_2^2-x_3^2\right)\Bigr],\label{first-chart-potIV-modelI-1}\\
 \frac{dx_2}{d\tau} 
&=&\frac{1}{2}x_2x_4\Bigl[6x_4-2\left(x_2^2+2x_3^2-1\right)+\alpha x_4^3 
 \nonumber \\ &&+\alpha x_4\left(1-x_1^2-x_2^2-x_3^2\right)\Bigr],\\
\label{first-chart-potIV-modelI-2}
\frac{dx_3}{d\tau} 
&=&\frac{1}{2}x_3x_4\Bigl[6x_4-2\left(x_2^2+2x_3^2-2\right)+\alpha x_4^3 
\nonumber  \\ &&+\alpha x_4\left(1-x_1^2-x_2^2-x_3^2\right)\Bigr],\\
\label{first-chart-potIV-modelI-3}
\frac{dx_4}{d\tau} &=& 
\frac{1}{2}x_4\Bigl[
3\left(1+x_1^2+x_2^2+x_3^2\right)-2x_4\left(x_2^2+2x_3^2\right)  \nonumber \\ 
&&+\alpha x_4^4+\alpha 
x_4^2\left(3+\left(1-x_2^2-x_2^2-x_3^2\right)\right)\Bigr].\label{
first-chart-potIV-modelI-4}
\end{eqnarray}
At $x_4=0$, the system behaves as   $\frac{dx_1}{d\tau}=0$, 
$\frac{dx_2}{d\tau}=0$, $\frac{dx_3}{d\tau}=0$ and $\frac{dx_4}{d\tau}=0$. This 
leads to the conclusion that the points at infinity  are 
$\left\{\left(x_1,x_2,x_3,0\right):x_1,x_2,x_3\in\mathbb{R}\right\}$. From the 
viewpoint of our phase-space domain, the critical points at infinity, located 
on the sphere $\mathbb{S}^4$, are    
$G^{+\infty}\Bigl\{\left(y_1,y_2,y_3,y_4,y_5\right)\in\mathbb{S}^4:y_5=0,y_4>0, 
y_3>0, y_2>0, y_1=\frac{1}{\sqrt{2}}\Bigr\}$. At the points 
$\left(x_1,x_2,x_3,0\right)$ three eigenvalues are zero, while one eigenvalue 
is positive, expressed as $\frac{3}{2}\left(1+x_1^2+x_2^2+x_3^2\right)$. This 
implies that these points exhibit unstable behavior.

In the second,  third and fourth charts, we again obtain   points  
$G^{+\infty}$ at infinity, which were identified in the first chart. Similarly, 
 the fifth chart results in a dynamical system similar to 
(\ref{reg-aut-sys-potIV-modelI-1})-(\ref{reg-aut-sys-potIV-modelI-4}), without 
offering any new solutions at infinity.

By using the re-parametrization $dN=-xd\tau$ in the phase space domain 
$\mathbb{D}_-=\mathbb{D}\setminus\{x\geq 0\}$, we write the system 
(\ref{aut-sys-potIV-modelI-1})-(\ref{aut-sys-potIV-modelI-4}) as
\begin{eqnarray}
    \frac{dx}{d\tau}&=&\frac{3}{2}x^2 \left(1+x^2+v^2+w^2+\xi^2\right) 
-xw^2-2x\xi^2 \nonumber \\ && 
+\frac{\alpha\left(1+x^2-v^2-w^2-\xi^2\right)}{2},\label{
n-reg-aut-sys-potIV-modelI-1
}  \\
\frac{dv}{d\tau}&=&-\frac{3}{2}xv\left(1-x^2-v^2-w^2-\xi^2\right),\label{n-reg-aut-sys-potIV-modelI-2}
\\ 
  \frac{dw}{d\tau}&=&-xw\left[x+\frac{3}{2}\left(1-x^2-v^2-w^2-\xi^2\right)\right],
  \label{n-reg-aut-sys-potIV-modelI-3}\\  
\frac{d\xi}{d\tau}&=&-x\xi\left[2x+\frac{3}{2}\left(1-x^2-v^2-w^2-\xi^2\right)\right].
\label{n-reg-aut-sys-potIV-modelI-4}
\end{eqnarray}
By applying the  same approach used in previous sections, we extract the 
critical points at infinity as
$G^{-\infty}\Bigl\{\left(y_1,y_2,y_3,y_4,y_5\right)\in\mathbb{S}^4:y_5=0,y_4>0,
y_3>0,y_2>0,y_1=-\frac{1}{\sqrt{2}}\Bigr\}$. This suggests that $G^{-\infty}$ 
also exhibits unstable behavior.

In summary, the original system  
(\ref{aut-sys-potIV-modelI-1})-(\ref{aut-sys-potIV-modelI-4}) has two points at 
infinity, namely, $G^{+\infty}$ and $G^{-\infty}$, both of which display 
unstable dynamics, and thus they cannot attract the Universe at late times.

\subsection{Model II: $Q=\Gamma \rho_m$}\label{appendix-8}

The system of equations 
(\ref{aut-sys-potIV-modelII-1})-(\ref{aut-sys-potIV-modelII-5}) is simplified by 
introducing the  re-parametrization $dN=x(1-z)d\tau$ for the 
phase space domain $\mathbb{D}_+=\mathbb{D}\setminus\{x\leq0\}$, as
\begin{eqnarray}
    \frac{dx}{d\tau}&=&(1-z)\Bigl[-\frac{3}{2}x^2 \left(1+x^2+v^2+w^2+\xi^2\right)+xw^2 \nonumber \\ &&+2x\xi^2\Bigr] -\frac{\alpha\left(1+x^2-v^2-w^2-\xi^2\right)}{2},\label{reg-aut-sys-potIV-modelII-1}  \\
\frac{dv}{d\tau}&=&\frac{3}{2}xv(1-z)\left(1-x^2-v^2-w^2-\xi^2\right),\label{reg-aut-sys-potIV-modelII-2}
\\ 
  \frac{dw}{d\tau}&=&xw\Bigl[x+\frac{3}{2}\Bigl(1-x^2-v^2-w^2 \nonumber \\ &&-\xi^2\Bigr)\Bigr](1-z),
  \label{reg-aut-sys-potIV-modelII-3}\\  
\frac{d\xi}{d\tau}&=&x\xi(1-z)\Bigl[2x+\frac{3}{2}\Bigl(1-x^2-v^2 \nonumber \\ &&-w^2-\xi^2\Bigr)\Bigr].
\label{reg-aut-sys-potIV-modelII-4}\\
\frac{dz}{d\tau}&=&\frac{3}{2}xz(1-z)^2\left(1-x^2-v^2-w^2-\xi^2\right).\label{reg-aut-sys-potIV-modelII-5}
\end{eqnarray}
The vector field  for the above system is defined by the components $P$, $Q$, 
$R$, $S$ and $T$, which are specified as: $P=(1-z)\Bigl[-\frac{3}{2}x^2 
\left(1+x^2+v^2+w^2+\xi^2\right)+xw^2+2x\xi^2\Bigr] 
-\frac{\alpha\left(1+x^2-v^2-w^2-\xi^2\right)}{2}$, 
$Q=\frac{3}{2}xv(1-z)\left(1-x^2-v^2-w^2-\xi^2\right)$, 
$R=xw\Bigl[x+\frac{3}{2}\Bigl(1-x^2-v^2-w^2-\xi^2\Bigr)\Bigr](1-z)$, 
$S=x\xi(1-z)\Bigl[2x+\frac{3}{2}\Bigl(1-x^2-v^2-w^2-\xi^2\Bigr)\Bigr]$ and 
$T=\frac{3}{2}xz(1-z)^2\left(1-x^2-v^2-w^2-\xi^2\right)$. We begin by projecting 
the vector field onto the upper five dimensional sphere $\mathbb{S}^{5+}$ using central 
projection, where the coordinates $y_1$, $y_2$, $y_3$, $y_4$, $y_5$, $y_6$ are 
related to $x$, $v$, $w$, $\xi$, $z$ through the following formulas: 
$y_1=\frac{x}{\Delta x}$, $y_2=\frac{v}{\Delta x}$, $y_3=\frac{w}{\Delta x}$, 
$y_4=\frac{\xi}{\Delta x}$, $y_5=\frac{z}{\Delta x}$ and $y_6=\frac{1}{\Delta 
x}$, where $\Delta x=\sqrt{1+x^2+v^2+w^2+\xi^2+z^2}$. The stability of the system is 
analyzed using six distinct local charts $\left(U_i,\phi_i\right)$, where 
$i=1,2,3,4,5,6$, and the sets $U_i$ are defined by  
$U_i=\left\{\left(y_1,y_2,y_3,y_4,y_5,y_6\right)\in\mathbb{S}^5:y_i>0\right\}$. 
The local map $\phi_i:U_i\longrightarrow\mathbb{R}^5$ is expressed as  
$\phi_i\left(y_1,y_2,y_3,y_4,y_5,y_6\right)=\left(\frac{y_n}{y_i},\frac{y_m}{y_i
},\frac{y_p}{y_i},\frac{y_q}{y_i},\frac{y_r}{y_i}\right)$, with the indices $n$, 
$m$, $p$, $q$, $r$ satisfying $n<m<p<q<r$, while $n, m, p, q, r\neq i$.

For the first chart,  with the selection 
$\phi_1\left(y_1,y_2,y_3,y_4,y_5,y_6\right)=(x_1, x_2, x_3, x_4, x_5)$, the dynamical system 
described by 
(\ref{reg-aut-sys-potIV-modelII-1})-(\ref{reg-aut-sys-potIV-modelII-5}) becomes
\begin{eqnarray}
\frac{dx_1}{d\tau}
&=&\frac{1}{2}x_5^2x_1\Bigl\{
2x_5\left(3x_5-x_2^2-2x_3^2\right)+x_4\Bigl\{2(x_2^2+2x_3^2) \nonumber \\ && 
+\gamma 
x_5^3-x_5\left[6+\gamma\left(x_1^2+x_2^2+x_3^2-1\right)\right]\Bigr\}\Bigr\},
\label{first-chart-potIV-modelII-1}\\
\frac{dx_2}{d\tau} 
&=&
\frac{1}{2}x_5^2x_2\Bigl\{
2x_5\left(1+3x_5-x_2^2-2x_3^2\right)\nonumber \\ && 
+x_4\Bigl\{2(x_2^2+2x_3^2-1) 
\nonumber \\ && +\gamma 
x_5^3-x_5\left[6+\gamma\left(x_1^2+x_2^2+x_3^2-1\right)\right]\Bigr\}\Bigr\}, 
\label{first-chart-potIV-modelII-2} 
\\
\frac{dx_3}{d\tau} 
&=&
\frac{1}{2}x_5^2x_3\Bigl\{2x_5\left(2+3x_5-x_2^2-2x_3^2\right) \nonumber \\ 
&& +x_4\Bigl\{2(x_2^2+2x_3^2-2) \nonumber \\ && +\gamma 
x_5^3-x_5\left[6+\gamma\left(x_1^2+x_2^2+x_3^2-1\right)\right]\Bigr\}\Bigr\},
\label{first-chart-potIV-modelII-3} \\
\frac{dx_4}{d\tau}
&=& 
\frac{1}{2}x_4\Bigl\{2x_5^3\left(3x_5-x_2^2-2x_3^2\right)+ 
\nonumber \\ && 3x_4^2\left(x_5^2-1-x_1^2-x_2^2-x_3^2\right)
\nonumber \\
&& 
+x_4x_5\Bigl\{3\left(1+x_1^2+x_2^2+x_3^2\right)+2x_5
\left(x_2^2+2x_3^2\right)\nonumber \\ && +\gamma x_5^4- 
x_5^2\left[9+\gamma\left(x_1^2+x_2^2+x_3^2-1\right)\right]\Bigr\}\Bigr\},
\label{first-chart-potIV-modelII-4}\\
\frac{dx_5}{d\tau} 
&=&
\frac{1}{2}x_5^2\Bigl\{
x_5\left(3x_5^2+3\left(1+x_1^2+x_2^2+x_3^2\right)\right.
\nonumber
\\ 
&&\left. 
-2x_5\left(x_2^2+2x_3^2\right)
\right)\nonumber \\ && 
+x_4\Bigl\{-3\left(1+x_1^2+x_2^2+x_3^2\right)+2x_5\left(x_2^2+2x_3^2\right)
\nonumber \\ &&  +\gamma x_5^4- 
x_5^2\left[3+\gamma\left(x_1^2+x_2^2+x_3^2-1\right)\right]\Bigr\}\Bigr\}.
\label{first-chart-potIV-modelII-5}
\end{eqnarray}
For $x_5=0$, the system evolves according to  the following differential 
equations: $\frac{dx_1}{d\tau}=0$, $\frac{dx_2}{d\tau}=0$, 
$\frac{dx_3}{d\tau}=0$, 
$\frac{dx_4}{d\tau}=-\frac{3}{2}x_4^3\left(1+x_1^2+x_2^2+x_3^2\right)$ and 
$\frac{dx_5}{d\tau}=0$. This implies that the points at infinity  
are   
$\left\{\left(x_1,x_2,x_3,0,0\right):x_1,x_2,x_3\in\mathbb{R}\right\}$. In the 
context of our phase-space domain, the critical points at infinity located on 
the sphere $\mathbb{S}^5$ are given by 
$J^{+\infty}\Bigl\{\left(y_1,y_2,y_3,y_4,y_5,y_6\right)\in\mathbb{S}^5:y_6=0,
y_5=0,y_4>0, y_3>0, y_2>0, y_1=\frac{1}{\sqrt{2}}\Bigr\}$. At the points 
$\left(x_1,x_2,x_3,0,0\right)$, all eigenvalues are zero, implying that the linear 
stability analysis does not provide conclusive results regarding stability. 
However, on the surface where $x_4=0$, the system follows the dynamics 
$\frac{dx_1}{d\tau}=x_1x_5^3\left(3x_5-x_2^2-2x_3^2\right)$, 
$\frac{dx_2}{d\tau}=x_2x_5^3\left(1+3x_5-x_2^2-2x_3^2\right)$, 
$\frac{dx_3}{d\tau}=x_3x_5^3\left(2+3x_5-x_2^2-2x_3^2\right)$, 
$\frac{dx_4}{d\tau}=0$ and 
$\frac{dx_5}{d\tau}=\frac{1}{2}
x_5^3\left(3x_5^2+3\left(1+x_1^2+x_2^2+x_3^2\right)-2x_5\left(x_2^2+2x_3^2\right
)\right)$, which suggest that near $x_5=0$, we have 
$\frac{dx_5}{d\tau}\cong\frac{3}{2}x_5^3\left(1+x_1^2+x_2^2+x_3^2\right)$, 
showing that $x_5$ is decreasing for $x_5<0$, and increasing for $x_5>0$. 
Therefore, $F^{+\infty}$ exhibits unstable behavior.

In the second, third and fourth charts, we find 
$J^{+\infty}$ (which has already been found as a critical point at infinity in the first chart) as the critical point at infinity. The fifth chart 
does not introduce any new critical points at infinity within our phase-space domain. 
The sixth chart leads to a dynamical system similar to 
(\ref{reg-aut-sys-potIV-modelII-1})-(\ref{reg-aut-sys-potIV-modelII-5}), without 
providing any new solutions at infinity, too.

In the phase-space domain $\mathbb{D}_-=\mathbb{D}\setminus\{x\geq 0\}$, the re-parametrization $dN=-x(1-z)d\tau$, transforms the system of equations 

(\ref{aut-sys-potIV-modelII-1})-(\ref{aut-sys-potIV-modelII-5}) as
\begin{eqnarray}
    \frac{dx}{d\tau}&=&(1-z)\Bigl[\frac{3}{2}x^2  
\left(1+x^2+v^2+w^2+\xi^2\right)-xw^2 \nonumber \\ &&-2x\xi^2\Bigr] 
+\frac{\alpha\left(1+x^2-v^2-w^2-\xi^2\right)}{2},
\label{n-reg-aut-sys-potIV-modelII-1} \end{eqnarray}
\begin{eqnarray}
\frac{dv}{d\tau}&=&-\frac{3}{2}xv(1-z)\left(1-x^2-v^2-w^2 
-\xi^2\right),\label{n-reg-aut-sys-potIV-modelII-2} \nonumber  \\ 
  \frac{dw}{d\tau}&=&-xw\Bigl[x+\frac{3}{2}\Bigl(1-x^2-v^2-w^2 \nonumber  \\ 
&&-\xi^2\Bigr)\Bigr](1-z),
  \label{n-reg-aut-sys-potIV-modelII-3}\\  
\frac{d\xi}{d\tau}&=&-x\xi(1-z)\Bigl[2x+\frac{3}{2}\Bigl(1-x^2-v^2 \nonumber  \\ 
&&-w^2-\xi^2\Bigr)\Bigr],
\label{n-reg-aut-sys-potIV-modelII-4}\\
\frac{dz}{d\tau}&=&-\frac{3}{2}xz(1-z)^2\left(1-x^2-v^2-w^2-\xi^2\right). 
\label{n-aut-sys-potIV-modelII-5}
\end{eqnarray}
Continuing with the same methodology as in previous sections,  the critical 
points at infinity are found to be: 
$J^{-\infty}\Bigl\{\left(y_1,y_2,y_3,y_4,y_5,y_6\right)\in\mathbb{S}^5:y_6=0,
y_5=0,y_4>0,y_3>0,y_2>0,y_1=-\frac{1}{\sqrt{2}}\Bigr\}$. This suggests that 
$J^{-\infty}$ also displays unstable behavior.

In summary, the original system 
(\ref{aut-sys-potIV-modelII-1})-(\ref{aut-sys-potIV-modelII-5}) has two points 
at infinity, namely $J^{+\infty}$ and $J^{-\infty}$, which are  both unstable, 
and therefore they cannot be the late-time solutions of the Universe.

\end{appendix}


\bibliography{biblio}
\end{document}